\documentclass[a4paper,11pt]
{scrartcl}

\usepackage[utf8]{inputenc}

\usepackage[english]{babel}

\usepackage{authblk,titling,url,amsfonts,amssymb,amstext,amsthm,amsmath,hyperref}
\usepackage[vlined,ruled,linesnumbered]{algorithm2e}
\usepackage{makecell}
\usepackage{tabu}
\usepackage{booktabs}

\usepackage{caption}

\usepackage{comment}

\usepackage{tikz}
\usepackage{tkz-graph}

\title{From Symmetry to Asymmetry:\\ Generalizing TSP Approximations by Parametrization}

\author{Lukas Behrendt\thanks{\texttt{
			\{lukas.behrendt,alexander.loeser,marcus.wilhelm\}@student.hpi.de}
		}}
\author{Katrin Casel\thanks{\texttt{
			\{katrin.casel, tobias.friedrich, gregor.lagodzinski\}@hpi.de}
		}}
\author{Tobias Friedrich\thanksmark{2}}
\author{\\ J.~A.~Gregor Lagodzinski\thanksmark{2}}
\author{Alexander Löser\thanksmark{1}}
\author{Marcus Wilhelm\thanksmark{1}}

\affil{\small Hasso Plattner Institute, University of Potsdam, Potsdam, Germany }
\date{}

\newcommand{\ezlkparam}{\ensuremath{z}}
\newcommand{\algoparam}{\ensuremath{k}}
\newcommand{\compcount}{\ensuremath{{k+1}}}

\newcommand{\asymfactor}{\ensuremath{\beta}}
\newcommand{\ezlk}{Generalized Christofides algorithm}
\newcommand{\avtwo}{Generalized tree doubling algorithm}
\newcommand{\metaM}{\ensuremath{M}}
\newcommand{\metam}{\ensuremath{v^M}}

\theoremstyle{plain}
\newtheorem{theorem}{Theorem}[section]
\newtheorem{corollary}[theorem]{Corollary}
\newtheorem{lemma}[theorem]{Lemma}
\newtheorem{proposition}[theorem]{Proposition}

\theoremstyle{definition}
\newtheorem{definition}[theorem]{Definition}

\begin{document}
\maketitle
\begin{abstract}
We generalize the tree doubling and Christofides algorithm, the two most common approximations for TSP, to parameterized approximations for ATSP. The parameters we consider for the respective parameterizations are upper bounded by the number of \emph{asymmetric distances} in the given instance, which yields algorithms to efficiently compute constant factor approximations also for moderately asymmetric TSP instances. As generalization of the Christofides algorithm, we derive a parameterized 2.5-approximation, where the parameter is the size of a vertex cover for the subgraph induced by the asymmetric edges. Our generalization of the tree doubling algorithm gives a parameterized 3-approximation, where the parameter is the number of asymmetric edges in a given minimum spanning arborescence. Both algorithms are also stated in the form of \emph{additive} lossy kernelizations, which allows to combine them with known polynomial time approximations for ATSP. Further, we combine them with a notion of symmetry relaxation which allows to trade approximation guarantee for runtime. We complement our results by experimental evaluations, which show 
that generalized tree-doubling frequently outperforms generalized Christofides with respect to parameter size.
\end{abstract}
\clearpage
\section{Introduction}
The famous traveling salesman problem asks for a shortest round trip through a  given set of cities. Its relation to the Hamiltonian cycle problem does not only show NP-hardness, but also implies that efficient approximation is not possible  for unrestricted instances, which is why the intercity distances are usually assumed to satisfy the triangle inequality. This restriction, usually called \emph{metric} traveling salesman problem, is one of the most extensively studied problems in combinatorial optimization, yet its approximability prevails as an open problem and active research area. 
Despite a recent breakthrough by Svensson et al.~\cite{constapx}, particularly the difference between symmetric and asymmetric distances remains rather poorly understood. In this paper we employ the tools of parameterized complexity as a new approach to explicitly study the effects of asymmetry on the approximability of the metric traveling salesman problem.
\subsection{Motivation}
Assuming symmetric distances, meaning that traveling from city A to city B requires the same cost as traveling from B to A, is certainly the most common restriction to the metric traveling salesman problem. In fact, it is so common that the name (metric) traveling salesman problem, TSP for short, is usually associated with this symmetric version, while the more general case is explicitly referred to as the \emph{asymmetric} traveling salesman problem, or just ATSP\@. Considering the known approximations for TSP and ATSP, symmetry seems to play a vital role. The currently best known approximation for TSP is the over 40 years old famous algorithm of Christofides~\cite{Christofides}, which guarantees a ratio of~$\tfrac 32$. For ATSP on the other hand, it was unclear for a long time if any constant factor approximation exists. 
From the $\log_2 n$-approximation by Frieze et al.~\cite{log_apx_82} to the subsequent improvements to a ratio of $0.999\log_2n$ by Bl{\"{a}}ser~\cite{apx_improved_2003a}, $0.842 \log_2 n$ by Kaplan et al.~\cite{apx_improved_2003}, $\frac{2}{3} \log_2 n$ by Feige and Singh~\cite{apx_improved_2007} and finally an asymptotic improvement to an $\mathcal{O}(\log n/\log\log n)$-approximation by Asadpour et al.~\cite{asadpour_o_2010}, ATSP proved to be a much more difficult problem to approximate. Significant effort went into the recent result of Svensson et al.~\cite{constapx} which gives the first constant factor approximation with a ratio of~5500, subsequently improved to~506 in~\cite{constapx_arxiv}. This still leaves a huge gap between the positive results for TSP and ATSP, while the currently known lower bounds of $\tfrac{123}{122}$ and $\tfrac{75}{74}$ for TSP and ATSP, respectively~\cite{inapx}, do not indicate such a vast difference in difficulty. This raises the question of how symmetry truly affects approximability.\par
Restriction to distances that satisfy the triangle inequality is a reasonable assumption in all scenarios where visiting cities more than once is acceptable. Finding a shortest tour that visits each city \emph{at least} once is sometimes called \emph{graphical} TSP~\cite{graphical_tsp} and translates to metric TSP by taking the shortest path metric, also called \emph{metric closure}. In comparison, restriction to symmetric distances seems less natural. Quite contrarily, there are scenarios where we expect asymmetry, for example in rush hour where traffic is low to leave the city while vehicles flood the streets that lead inside (or vice versa). Such phenomena can result in unbounded violations of symmetry while the triangle inequality remains satisfied. However, even without traffic, restricted access such as road blocks or one-way streets may result in asymmetry. A recent study by Mart{\'{\i}}nez Mori and Samaranayake~\cite{bd-asymmetry} shows that road networks exhibit asymmetry even when only the lengths of the shortest paths are considered. Their investigations however also reveal that most asymmetries are insignificantly small.\par
With these few but existing significant asymmetries in mind, we consider algorithms that are not purely polynomial but may spend exponential time with respect to some measure of the degree of asymmetry. Our basic objective is to salvage the approximability of TSP for ATSP by allowing this increase in runtime. In particular, we consider the 1.5-approximation of Christofides and the 2-approximation derived by tree doubling, and generalize these with respect to parameters that describe the effect of asymmetry on these algorithms.
  Formally, our algorithms fall into the framework of parameterized approximations (see for example the survey of Marx~\cite{param_apx}), which means that they guarantee a fixed performance ratio and exhibit a runtime of the form $poly(n)f(k)$, where $f$ is an arbitrary function, $n$ is the size of the instance and $k$ is a measure for asymmetry. 
  This parameterized approach aims to offer efficient algorithms for instances of low asymmetry and to improve our understanding of the challenges asymmetric distances pose to the design of efficient approximation algorithms.
\subsection{Related Work}\label{sec:relwork}
Conceptually, our approach to generalize approximations for TSP can be seen as a study of \emph{stability} with respect to asymmetry in the framework of \emph{stability of approximation} by B{\"{o}}ckenhauer et al.~\cite{stability}. Probably the most extensively studied  stability measure for (A)TSP is the $\beta$-triangle inequality, also called \emph{parameterized} triangle inequality, which refers to the requirement $c(u,v)\leq \beta(c(u,w)+c(w,v))$ for all $u,v,w\in V$ with $u\not=v\not=w$. 
For ATSP with $\beta$-triangle inequality, the $\frac{1}{2(1-\beta)}$-approximation derived by Kowalik and Mucha~\cite{sharpened_atsp_09}  for  $\beta\in (\frac 12,1)$ improves upon a series on previous results~\cite{sharpened_atsp,sharpened_atsp_03,sharpened_atsp_08} and is also known to be tight with respect to the cycle cover relaxation as lower bound. For TSP, a recent survey of Klasing and M{\"{o}}mke~\cite{new_stability_survey} gives a summary of the known results with $\beta$-triangle inequality.\par
%
Regarding a measure for asymmetry of ATSP, Mart{\'{\i}}nez Mori and Samaranayake~\cite{bd-asymmetry}  define  the \emph{asymmetry factor}~$\Delta$ as the maximum ratio between the length of the shortest paths from $A$ to $B$ and  $B$ to $A$ over all cities $A, B$. They show that the Christofides algorithm is~$\frac 32$-stable with respect to this measure, meaning that Christofides can be used to compute a $\frac 32\Delta$-approximation for instances  with asymmetry factor bounded by $\Delta$. We discuss combining this asymmetry factor with our results but our focus is on parameterized approximation, meaning that the  asymmetry measures regulate the runtime and not the performance ratio.\par
So far, there are only a few parameterized approximations for (variations of) TSP\@. Marx et al.~\cite{pathwidth_atsp} consider ATSP on a restricted graph class called $k$-nearly-embeddable. They derive approximations where the ratio and the runtime depend on structural parameters of the given instance. A true parameterized approximation for a TSP type problem is given by B{\"{o}}ckenhauer et al.~in~\cite{deadline} for deadline TSP, a generalization of TSP where some cities have to be reached by the tour within a given deadline. They give a 2.5-approximation that requires exponential time only with respect to the number of cities with deadline.\par
%
Another interesting approach to invest moderate exponential time to achieve better performance ratios for ATSP is given by Bonnet et al.~in~\cite{atsp_time_ratio_tradeoff}. They derive what could be called a subexponential approximation scheme, more precisely, a routine that allows to compute for any $r\leq n$ a $\log r$-approximation for ATSP that requires time $\mathcal O^*(2^{\frac nr})$.
\subsection{Our Results}
We derive parameterized approximations based on generalizations of the Christofides and the tree doubling algorithm, and choose for each generalization a suitable parameter. Both parameters under study are upper bounded by the number of \emph{asymmetric distances} in the given instance, meaning the pairs of vertices $u$ and $v$ for which the cost of traveling from $u$ to $v$ is cheaper than the cost of traveling from $v$ to $u$. As generalization of the Christofides algorithm, we derive in Section~\ref{sec:ezlk} a parameterized 2.5-approximation, where the parameter is the size of a vertex cover for the subgraph induced by the asymmetric edges (the pair of edges corresponding to an asymmetric distance) in the instance graph. 
Our main result is the more elaborate generalized tree doubling algorithm which uses parameterization by the number of asymmetric edges in a given minimum spanning arborescence. The resulting parameterized 3-approximation is presented in Section~\ref{sec:algo_v2}.
These two results  are also stated in the form of \emph{additive} lossy kernelizations, which allows for a combination with known polynomial time approximations for ATSP\@. Further, we consider combining them with the asymmetry factor $\Delta$  by Mart{\'{\i}}nez Mori and Samaranayake~\cite{bd-asymmetry}. In Section~\ref{sec:beta} we show that this allows to trade approximation ratio for runtime as follows. We adjust our parameterized approximations to treat distances with asymmetry factor $\Delta\leq \beta$ for some $\beta\geq 1$ as symmetric, which shrinks both parameters to consider only the more severe asymmetries.
 For generalized Christofides, this relaxation yields a ratio of~$\frac74+\frac 34\beta$, and for generalized tree doubling a ratio of~$2+\beta$, each parameterized by their respective parameters for the severe asymmetries. Since the chosen parameters are theoretically incomparable, 
 we conducted experiments on the ATSP instances of the TSPLIB~\cite{reinelt_tsplibtraveling_1991}. Section~\ref{sec:experiments} presents the results which show that the approximation ratio remains below~2 for both algorithms, even when some relaxation with the asymmetry factor is used to diminish the kernel size. Further, we observe that the generalized tree doubling algorithm frequently outperforms generalized Christofides with respect to parameter size. 
  Due to space restrictions, some proofs are moved to Appendix~\ref{app:lemmata}.

\section{Preliminaries}
Throughout the paper, instances of ATSP are always simple complete directed graphs denoted by~$G=(V,E,c)$ with non-negative cost function $c$ on~$E$. For $u,v\in V$, $(u,v)$ denotes the edge from $u$ to $v$ and $c(u,v)$ denotes its cost. If the cost function $c$ satisfies the triangle inequality, i.e., $c(u,v) + c(v,w) \ge c(u,w)$ for all $u,v,w\in V$, we call $G$ \emph{metric}. If the graph is not clear from context, we use $V[G]$ and $E[G]$ to denote the vertices and edges of $G$, respectively.\par
For a not necessarily complete graph $G'$, a \emph{trail} is a sequence of vertices of $G'$ in which each vertex is equal to or adjacent to its successor. We use the term \emph{path} for a trail containing no vertex twice.
We say a \emph{circuit} is a trail where the last vertex is connected to the first vertex.
If a circuit visits no vertex twice, we call it a \emph{cycle}.
We denote a trail by $v_1, \dots, v_n$ and a circuit by $(v_1, \dots, v_n)$. A \emph{tour} of $G'$ is a cycle that visits each vertex of $G'$.\par
If $G$ is a metric ATSP instance, every trail can be turned into a path visiting the same vertices via a \emph{metric shortcut} without increasing the cost. A metric shortcut is constructed by removing multiple occurrences of each vertex in the path.
All tours in $G$ are valid solutions, and we use $c^*(G)$ to denote the cost of an optimal solution for $G$.\par
For an ATSP instance $G=(V,E,c)$ and a subset of vertices $V' \subseteq V$, we denote the vertex-induced subgraph by $G[V']$; for a set of edges $E' \subseteq E$, we denote the edge-induced subgraph by $G[E']$. Slightly abusing notation, $G[V']$ and  $G[E']$ also inherit the edge-weights of $G$. Further, for a subgraph $G'$ of $G$, we use $c(G')$ to denote the sum of all edge costs in $G'$.  
The following results on subgraphs are used in later sections.
\begin{lemma}\label{lemma:induced_graph_metric}
    Let $G$ be a metric graph and $V' \subseteq V$.
    Then, $G[V']$ is metric as well.
\end{lemma}
\begin{lemma}\label{lemma:induced_graph_has_leq_tour}
    Let $G$ be a metric ATSP instance and $V' \subseteq V$. Then, $c^*(G[V']) \leq c^*(G)$.
\end{lemma}
We also use one other transformation for ATSP instances which we refer to as  \emph{minors}.
In our context, $G'$ is a \emph{minor} of $G$ if there is a series of contractions which, starting from $G$, result in $G'$. A \emph{contraction} of an edge $(u,v)$ is an operation that replaces the vertices $u$ and $v$ with a single vertex $\mathit{uv}$ and sets $c(w,uv)=\min\{c(w,u), c(w,v)\}$ and $c(uv,w)=\min\{c(u,w), c(v,w)\}$ for all $w\in V\setminus\{u,v\}$.\par
We consider ATSP with parameterizations which gives instances of a \emph{parameterized optimization problem}. Instances are of the form $(G,k)$, where $G$ is an edge-weighted graph and $k$ is a given value in $\mathbb N$ for which we allow increase in runtime.  Our parameterized approximations are stated with explicit runtimes instead of such formal parameterization. We however also use \emph{lossy kernelization} as defined by Lokshtanov et al.~\cite{lossykernel} which formalizes preprocessing for approximation algorithms. As size measure for $G$ we consider the number of vertices $|G|=|V|$. For $\alpha \in \mathbb{R}$ with  $\alpha\ge 1$, a \emph{linear $\alpha$-approximate kernelization} for a parameterized optimization problem $\Pi$ is a pair of polynomial time algorithms  such that:
    \begin{itemize}
        \item The \emph{kernelization algorithm} takes an instance $(I, k)$ of $\Pi$ as input and returns as \emph{kernel} an instance $(I', k')$ of $\Pi$ such that $|I'| + k'$ is upper bounded by a \emph{linear} function of $k$. 
        \item The \emph{solution lifting algorithm} returns for an instance $(I, k)$ of $\Pi$, its kernel $(I', k')$ and a $\gamma$-approximate solution $s'$ for  $(I', k')$ as input, a $(\gamma \cdot \alpha)$-approximate solution for $(I, k)$.
    \end{itemize}
This definition assumes that the approximation error introduced by the kernelization scales linearly with the ratio~$\gamma$ with which the instance $(I', k')$ is solved.
This has the drawback that if the kernelization only adds a constant approximation error, the kernelization cannot be accurately described. In order to amend this, we also define an \emph{linear $\alpha$-additive kernelization} with the difference that the solution lifting algorithm returns a $(\gamma + \alpha)$-approximate solution instead of a $(\gamma \cdot \alpha)$-approximate solution.
Note that every $\alpha$-additive lossy kernelization is also a $(1 + \alpha)$-approximate lossy kernelization, because $\gamma \ge 1$ and thus $\gamma+ \alpha \le (1 + \alpha)\gamma$.

\section{Generalized Christofides Algorithm}\label{sec:ezlk}
The Christofides algorithm~\cite{Christofides} is the best known polynomial approximation for TSP with performance ratio~1.5. It builds an approximate solution for an instance $G$ by first computing a minimum spanning tree~$T$ for $G$ and then adds to this tree a minimum cost perfect matching~$M$ on the vertices of odd degree in $T$. The resulting subgraph has even degree, so it is possible to compute an Eulerian cycle for it, which results in a circuit of cost $c(T)+c(M)$ that visits all vertices. Metric shortcuts turn this circuit into a tour. Since taking every second edge in an optimal tour gives a perfect matching of the whole graph, metric shortcuts can be used to show that the edges in~$M$  have a cost of at most $\frac 12 c^*(G)$, and a minimum spanning tree also has a cost of at most $c^*(G)$, which yields the 1.5-approximation.\par
Regarding ATSP, the most dire problem of this approach is that combining the two edge sets from $T$ and $M$ to an Eulerian circuit is impossible if some edges are directed in the wrong direction, and it is unclear how to restrict $T$ and $M$ accordingly while keeping the relation of their cost to the optimum value. Due to this conceptual problem, our approach reduces an ATSP instance to a TSP instance for which the Christofides algorithm can be applied. Observe that such a reduction cannot simply be designed by brute-force guessing the correct set of directed edges in an optimal solution; fixing a subset of directed edges to be in a solution can not be modeled as an undirected instance (unless the instance is completely directed, in which case this approach results in a brute-force guess of the whole tour). 
The design of our  algorithm is instead based on a simple structural insight that allows the use of the Christofides algorithm on a symmetric subgraph.\par
We first explain an easier variant of the algorithm in the form of a lossy kernelization before improving it further. The idea is to divide the graph into two overlapping subgraphs: an asymmetric subgraph, which is returned as the kernel, and a symmetric subgraph. Formally, for an instance $G$ of ATSP with cost function $c$ we define the set of \emph{asymmetric edges} by $E_a=\{(u,v),(v,u)\mid u,v\in V[G], c(u,v)\not=c(v,u)\}$ and the set of \emph{asymmetric} and \emph{symmetric vertices} by $V_a=\{v\in V[G]\mid (v,u)\in E_a \text{ for some } u\in V\}$, and $V_s=V[G]\setminus V_a$, respectively.\par 
The kernelization algorithm returns $G[V_a \cup \{ v \}]$ as the kernel, where $v$ is an arbitrary vertex in $V_s$. The solution lifting algorithm receives a tour of $G[V_a \cup \{ v \}]$ as input, and computes a 1.5-approximate tour of $G[V_s]$ via the Christofides algorithm.
This way, the two tours visit every vertex in the graph and overlap in $v$.
They can therefore be merged into a circuit that visits the entire graph without additional cost by gluing them together at the common vertex~$v$ before employing metric shortcuts to obtain a tour. Overall, this approach gives a 1.5-additive kernelization parameterized by $|V_a|$. With an exponential time exact solution of the kernel, this yields a parameterized~2.5-approximation with parameter $|V_a|$.\par
To improve this approach, consider a vertex cover $\mathit{VC}$ of $G[E_a]$.
The complement of $\mathit{VC}$ forms an independent set in $G[E_a]$, implying that $G$ contains no asymmetric edges between vertices in $V_s\cup (V_a \setminus \mathit{VC})$.
This can be exploited to improve the lossy kernelization to a structural parameter $\ezlkparam$ being the size of a vertex cover in $G[E_a]$. The improved algorithm uses a vertex cover $\mathit{VC}$ in $G[E_a]$, selects a vertex $v \in V_s$ and returns $G[\mathit{VC} \cup \{ v \}]$ as kernel and lifts the solution like before; observe that $G[V[G]\setminus \mathit{VC}]$ is a symmetric graph which allows to use the Christofides algorithm to derive a 1.5-approximation for it.\par 
There are multiple ways of finding a vertex cover in $G[E_a]$. One option is to use one of the polynomial 2-approximations. 
Another option resulting in a smallest possible kernel is to compute a minimum vertex cover by known FPT algorithms, for example the $\mathcal O(1.2738^{\ezlkparam})$-algorithm by Chen et al.~\cite{CheKanXia2010}, which results in a parameterized approximation but is technically not a lossy kernelization. Therefore, we will state our result with parameter $\ezlkparam$ being the size of a given vertex cover for $G[E_a]$. This algorithmic approach yields:
\begin{theorem}\label{thm:1.5kernel}
    Metric ATSP admits a linear $1.5$-additive lossy kernelization for parameter $\ezlkparam$, size of a vertex cover of the subgraph induced by all asymmetric edges.
\end{theorem}
Using the $\mathcal O(1.2738^{\ezlkparam})$-algorithm by Chen et al.~\cite{CheKanXia2010} to provide a minimal vertex cover for Theorem~\ref{thm:1.5kernel}, and the dynamic programming algorithm by Held and Karp to solve ATSP for the kernel exactly~\cite{held_dynamic_1962} with a runtime in $\mathcal O(2^{\ezlkparam}\ezlkparam^2)$ yields the following.
\begin{corollary}\label{cor:2.5apx}
Metric ATSP can be $2.5$-approximated in $\mathcal{O}(n^3 + 2^{\ezlkparam}\ezlkparam^2$), where $\ezlkparam$ is the size of a minimum vertex cover of the subgraph induced by all asymmetric edges.
\end{corollary}
Instead of solving the kernel with an exact algorithm, its solution can also be approximated.
Using the lossy kernelization with a 2-approximate vertex cover and solving the kernel with the $\frac{2}{3} \log n$-approximation of Feige and Singh~\cite{apx_improved_2007} yields the following interesting result.
\begin{corollary}\label{cor:poly_apx}
Metric ATSP can be $(\frac{2}{3} \log x + \frac{3}{2})$-approximated  in polynomial time, where $x= \min\left( 2\ezlkparam+1, |V_a| \right)$, $V_a$ is the set of asymmetric vertices and $\ezlkparam$ is the size of a minimum vertex cover for the subgraph induced by all asymmetric edges.
\end{corollary}
This improves upon the approximation ratio of $\frac{2}{3} \log n$ if  $\frac{x}{n} < 2^{-\frac{9}{4}}$, meaning that the kernel only contains a sufficiently small fraction of the vertices. We note that the result of Asadapour et al.~\cite{asadpour_o_2010} gives a polynomial $(8\log(z)/\log\log(z)+\tfrac 32)$-approximation, which is asymptotically stronger but less suitable for the instances with small values of $z$ we are interested in.\par
It remains to see if this approach can be improved further. Aiming for a smaller kernel than a vertex cover for $G[E_a]$ seems difficult as this results in asymmetric edges in the reduced graph. Regarding a possible improvement of the ratio, one might hope to salvage the ratio of 1.5 for TSP, obtained by the Christofides algorithm, for ATSP. However, the additive ratio of 1.5 in Theorem~\ref{thm:1.5kernel} is asymptotically tight, which can be shown as follows. \par
\begin{figure}
    \centering
    \includegraphics[page=4,width=0.35\textwidth,trim={4cm, 6cm, 4cm, 5.5cm},clip]{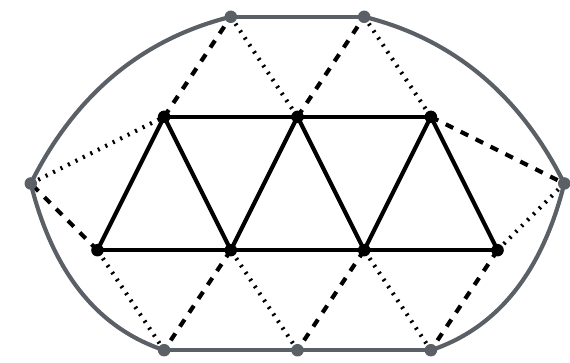}
    \caption{The constructed graph $G_k$ for $k=7$.
    Black and gray edges are symmetric with cost 2.
    Dotted edges are symmetric with cost 1.
    Dashed edges are asymmetric, with cost 1 from a gray to a black vertex and cost 2 from a black to a gray vertex. Metric closure gives the ATSP instance.}
    \label{fig:2.5tight_figure1}
\end{figure}
We define a family of graphs $G_k$ for $k \in \mathbb{N}, \, k>2$ such that the approximation ratio converges to 2.5 for increasing $k$.
Depending on whether the vertex cover is approximated or computed exactly, the construction of the graphs differs slightly.  Figure~\ref{fig:2.5tight_figure1} describes $G_7$ for the version of the algorithm in which the vertex cover is solved exactly.
The black zig-zag pattern is the textbook example for the tightness of the Christofides algorithm.
Consequently, the idea is that a possible minimal vertex cover consists of all vertices of the gray cycle, such that the black zig-zag pattern becomes the symmetric instance to be solved with the Christofides algorithm.
The gray cycle is the kernel and solving it exactly yields a tour of cost $2k$, once around the circle.
Together with the approximation on the symmetric subgraph, which converges to $3k$, this results in a tour of length $5k$.
As the optimal tour takes the dotted and dashed edges in the cheaper direction and has cost $2k$, we deduce that 2.5 is asymptotically tight for Corollary~\ref{cor:2.5apx}. For a full proof of this construction see Section~\ref{sec:2.5_tightness}.
\section{Generalized Tree Doubling Algorithm}\label{sec:algo_v2}
One widely known constant factor approximation for TSP is the tree doubling algorithm, which
computes a minimum spanning tree and doubles every edge in it to ensure the existence of an Eulerian circuit.
Since the circuit uses every original edge exactly twice, it is twice as expensive as the tree, which itself is at most as expensive as the optimum tour.
Thus, by transforming the circuit with metric shortcuts, a 2-approximate tour is found.
In order to adapt this approach to ATSP we employ a \emph{minimum spanning arborescence} (MSA) as the directed variant of a minimum spanning tree. However, the previous approach fails when doubling every edge, as the cost of reversed edges of the arborescence can be arbitrarily higher than the direction contained in the arborescence. These edges are the core of the problem and hence the basis for our parameterization to generalize the tree doubling algorithm.
\par
Formally, we define the notion of a \emph{one-way edge} in $G$ as an edge $(u,v) \in E[G]$  for which $c(u,v) < c(v,u)$. Given a spanning arborescence $A$ for $G$, we define the parameter~$\algoparam$ for our parameterized approach to be the number of one-way edges in $A$. Note that an MSA is easy to compute, e.g.\ with the Chu--Liu/Edmonds algorithm~\cite{edmonds_optimum_1967}. Still, we give it as input to our algorithm, similar to the vertex cover for the generalization of the Christofides algorithm. The reason for this is that a graph can contain an exponential number of MSAs, which can have different numbers of one-way edges, and it is unclear if it is possible to efficiently compute an MSA with a minimum number of one-way edges. In a nutshell, our  algorithm removes all one-way edges from the given arborescence, computes a tour for each resulting connected component by an altered tree doubling routine and uses exponential time in $\algoparam$ to compute a meta-tour to connect these subtours to a solution for the whole graph.
\par
Let $A$ be an MSA for $G$ and let $T_1, \dots, T_\compcount$ be the connected components in the graph created by deleting all one-way edges from $A$. We construct the meta-graph $\metaM$ by contracting each set of vertices $V[T_i]$ corresponding to the connected components, to one vertex $v_i^M$ with our notion of contraction to a minor. This results in $V[\metaM]=\{\metam_1,\dots,\metam_\compcount\}$ and  $c(\metam_i, \metam_j) = \min\left(\left\{ c(t_i, t_j) \mid t_i \in V[T_i], t_j \in V[T_j] \right\}\right)$ for all $\metam_i, \metam_j \in V[M]$ with $i\neq j$. 
\begin{lemma}\label{lemma:tspm_leq_tspg}
    Let $G$ be a metric ATSP instance and let $\metaM$ be a minor of $G$. Then,  \mbox{$c^*(\metaM) \le c^*(G)$}.
\end{lemma}
\begin{figure}[t]
    \centering
    \includegraphics[page=1,width=0.28\textwidth]{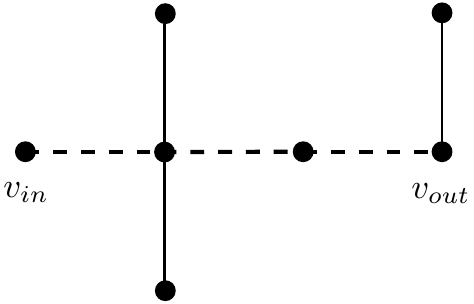}
    \hspace{.3cm}
    \includegraphics[page=2,width=0.28\textwidth]{euler}
    \hspace{.3cm}
    \includegraphics[page=3,width=0.28\textwidth]{euler}
    \caption{Exemplary constructive route for a suitable path through a component. The first image depicts the spanning tree of the component, with $P_i$ highlighted as dashed. The second depicts the trail through the partially doubled edges. The third depicts the resulting path.} 
    \label{fig:example_hamilton_path}
\end{figure}
Lemma~\ref{lemma:tspm_leq_tspg} implies that $c^*(\metaM)\leq c^*(G)$.
 Since  $\metaM$ contains  $\compcount$ vertices, we can compute an optimal tour $\tau'$ in time exponential only in our chosen parameter.
It remains to explain how to turn $\tau'$ into a tour of $G$ in the next step. Consider a vertex $\metam_i$ in $M$  (which corresponds to the component $T_i$) and assume without loss of generality that in $\tau'$ it is preceded by $\metam_{i-1}$ and precedes $\metam_{i+1}$.
Further, let $(v^{T_{i-1}}_{out}, v^{T_i}_{in})$ be the cheapest edge between $T_{i-1}$ and $T_i$, and let $(v^{T_{i}}_{out}, v^{T_{i+1}}_{in})$ be the cheapest edge between $T_{i}$ and $T_{i+1}$.
The goal is then to find a path $\chi_i$ that starts in $v^{T_i}_{in}$, ends in $v^{T_i}_{out}$, and spans all vertices in $T_i$. Replacing $v^{T_i}$ in $\tau'$ by $\chi_i$ for each $i$ turns $\tau'$ into a tour for $G$. However, the cost of $\chi_i$ has to be reasonably bounded.
\par
Such a path $\chi_i$ through $T_i$ can be found by adapting the tree doubling algorithm. We treat $T_i$ undirected  and double in it all edges that are not on the shortest path $P_i$ from~$v^{T_i}_{in}$ to $v^{T_i}_{out}$. The resulting graph contains an Eulerian trail from $v^{T_i}_{in}$ to $v^{T_i}_{out}$, which is turned into a  path by  metric shortcuts ensuring that $v^{T_i}_{in}$ and $v^{T_i}_{out}$ remain start and end node, see  Figure~\ref{fig:example_hamilton_path} for an example. Section~\ref{sec:algo_v2_sub} gives a detailed description of this adapted tree doubling. 
\par
For the cost of $\chi_i$ note that it contains for each edge $(u,v)$ in $T_i$ at most both $(u,v)$ and $(v,u)$. Since there are no one-way edges in $T_i$, any reverse edge is at most as expensive as the direction present in $T_i$. Consequently, the cost of $\chi_i$ is at most twice the cost of the edges in $T_i$ and the sum of all $\chi_i$ is at most $2c^*(G)$. In combination with the cost of at most $c^*(G)$ for the meta-tour, this yields a 3-approximate solution given by the following formal algorithm.
\begin{algorithm}
\caption{\avtwo}\label{algo_v2_main}
    \DontPrintSemicolon
    $T_1,\dots,T_\compcount \gets$ components of $A[\{e \in E[A] \mid e$ is not a one-way edge$ \}]$\;
    $\metaM \gets$ complete graph with vertices $\metam_1, \dots, \metam_\compcount$\;
    \ForEach{$\metam_i, \metam_j \in V[M]$ with $i \neq j$}{
        $c(\metam_i, \metam_j) \gets \min\left(\left\{ c(t_i, t_j) \mid t_i \in V[T_i], t_j \in V[T_j] \right\}\right)$\;
    }
    $\tau' \gets$ an optimum tour of $M$\;\label{algo:line:solve_M}
    \ForEach{subsequent $\metam_{i-1}, \metam_i, \metam_{i+1} \in \tau'$}{
        $v^{T_{i-1}}_{out}, v^{T_i}_{in} \gets \mathrm{argmin}_{t_{i-1} \in V[T_{i-1}], t_i \in V[T_i]}(c(t_{i-1}, t_i))$\;
        $v^{T_i}_{out}, v^{T_{i+1}}_{in} \gets \mathrm{argmin}_{t_i \in V[T_i], t_{i+1} \in V[T_{i+1}]}(c(t_i, t_{i+1}))$\;
        $\chi_i \gets $ path from $v^{T_i}_{in}$ to $v^{T_i}_{out}$ spanning $T_i$  by adjusted tree doubling\;
    }
    $\tau \gets$ concatenation of $(\chi_1,\dots,\chi_\compcount)$\;
    \Return $\tau$
\end{algorithm}
\begin{theorem}\label{thm:3_approx}
    A 3-approximation for metric ATSP can be computed in time $\mathcal{O}(2^\algoparam \algoparam^2 + n^2)$, where $\algoparam$ is the number of one-way edges in a given minimum spanning arborescence.
\end{theorem}

Contrary to the algorithmic approaches described in Section~\ref{sec:ezlk}, we did not present the above approach as a lossy kernelization.
The reason is that the constructed meta-graph $\metaM$ is not metric and thus no actual kernel.
For evidence, recall that $\metaM$ is a minor of the input graph $G$ and note that contractions do not preserve the triangle inequality.
Still, there could be a better way than to solve $\metaM$ with the expensive Held-Karp dynamic program, as the graph was obtained by contracting a metric graph and might have structural properties that can be exploited.
However, it can be shown that any non-metric graph $G$ is a minor of some metric graph $\hat{G}$ by a construction indicated in  Figure~\ref{fig:always_minor_of_metric_short}.  Include in $\hat{G}$ edges $(u_v, v_u)$ and $(v_u, u_v)$ for each pair of vertices $u,v \in V[G]$ and assign them the costs $c(u, v)$ and $c(v, u)$, respectively.
        All missing edges are then assigned a cost higher than the highest edge cost in $G$.
        This graph is metric and by contracting all edges between vertices $V_u=\{u_v \mid v\in V[G], v\neq u\}$ for each $u\in V[G]$, $G$ emerges as a minor of $\hat{G}$. This yields:
        \begin{figure}
    \centering
    \includegraphics[]{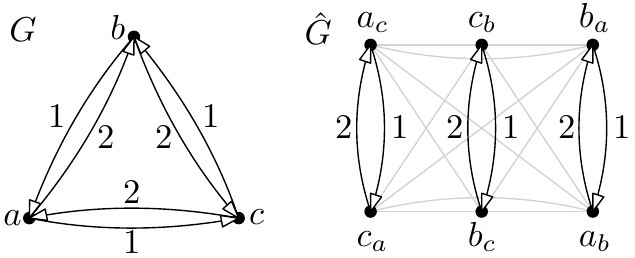}
    \caption{The construction for Proposition~\ref{prop:always_minor_of_metric_short}. The gray edges in $\hat{G}$ are assigned a cost higher than any other edge.}
    \label{fig:always_minor_of_metric_short}
\end{figure}
\begin{proposition}\label{prop:always_minor_of_metric_short}
    Let $G$ be a complete, directed graph with cost function $c$.
    Then, there exists a complete, metric graph $\hat{G}$ of which $G$ is a minor.
\end{proposition}
Computing a tour for the graph $\metaM$ is however related to the so-called \emph{generalized traveling salesman problem} (GTSP), which can be tracked back to publications of Henry-Labord{\`e}re and Saksena~\cite{henry-labordere:69,
saksena:70}. Given a partition  of the cities into $r$ sets, GTSP asks to find a minimum cost tour that contains (at least) one vertex from each of the $r$ sets. Unfortunately, there are no known efficient solutions to solve or approximate GTSP. However, we observe that using an optimal GTSP tour for the vertex sets corresponding to $T_1,\dots,T_{\compcount}$ instead of the tour through the constructed minor $\metaM$, still yields a 3-approximate solution. In fact, this remains true even if we fix one arbitrary city for each set, which yields a subgraph $\metaM'$ that is just an induced subgraph and hence a metric ATSP instance. As a result of this and by replacing each vertex in a tour of $\metaM'$ by a cycle built by using tree doubling for each $T_i$ we obtain the following.
\begin{theorem}\label{thm:3kernel}
    Metric ATSP admits a linear $2$-additive lossy kernelization with respect to  parameter \algoparam, number of one-way edges in a minimum spanning arborescence.
\end{theorem}
Aside from the fact that we were unable to construct a tight example for Theorem~\ref{thm:3_approx}, observing that the choice of any arbitrary vertex as representative---as done for Theorem~\ref{thm:3kernel}---still yields a 3-approximation causes us to conjecture that our more sophisticated generalization of the tree doubling algorithm has in fact a performance ratio of~2. The problem is that proving such a ratio requires an exploitable connection between the cost for the paths $\chi_i$ and the cost for the meta-tour through $\metaM$.
\par
\begin{figure}
    \centering
    \includegraphics[width=0.6\textwidth]{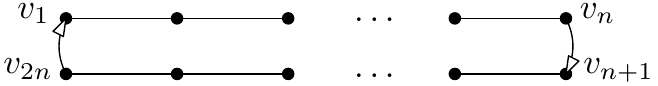}
    \caption{The constructed graph that shows the tightness of the algorithm in Theorem~\ref{thm:3kernel}.}
    \label{fig:3_tightness}
\end{figure}

For Theorem~\ref{thm:3kernel}, the estimated ratio is indeed asymptotically tight. This can be seen by the example illustrated in Figure~\ref{fig:3_tightness}. Consider for a fixed $n\in \mathbb N$ a cycle on $2n$ vertices where only the first and the $(n+1)$st edge is directed. Assign a cost of 1 to all edges of this cycle. Let $G$ be the metric closure of this graph and let $T$ be an MSA for $G$ that only contains one of the directed edges; observe that any MSA for $G$ has to be build from $2n-1$ out of the $2n$ edges from the initial cycle that yields all costs in $G$. This results in two components, the path from $v_1$ to $v_n$  and the path from $v_{n+1}$ to $v_{2n}$. Pick $v_n$ and $v_{2n}$ as representatives to build the kernel. Trivially, $(v_n,v_{2n})$ is an optimal solution for the kernel and the lifting algorithm replaces $v_n$ by the subtour $S_1=(v_n, v_{n-1},\dots, v_1)$, and $v_{2n}$ by the subtour $S_2=(v_{2n},v_{2n-1}\dots,v_{n+1})$, which overall results in a solution of cost $c(S_1)+c(v_1,v_{2n})+c(S_2)+c(v_{n+1},v_n)=(n-1)+(2n-1)+(n-1)+(2n-1)=6n-4$.
Since an optimal solution has a cost of $2n$, this shows that the ratio of~3 in Theorem~\ref{thm:3kernel} is asymptotically tight.

\section{Trading Approximation Quality for Runtime}\label{sec:beta}
In real life applications, we expect graphs to contain many small asymmetries. These lead to a relatively big kernel size, but do not have a big impact overall.
Therefore, ignoring asymmetric edges where both directions have similar cost and trading some approximation quality for a smaller kernel yields an intriguing perspective on the problem. As a formal way to describe this moderate asymmetry, we use the asymmetry factor of Mart{\'{\i}}nez Mori and Samaranayake~\cite{bd-asymmetry} as already introduced in Section~\ref{sec:relwork}. Since $\Delta$ is usually used for the maximum degree, and we want to describe variable restrictions of the asymmetry factor, we use the character~$\beta$ to describe moderate asymmetry.
    For $\beta\geq 1$ we call an edge $(u,v)$ \emph{\asymfactor-symmetric} if $\frac{1}{\asymfactor} \le \frac{c(u,v)}{c(v,u)} \le \asymfactor$, otherwise the edge is called \asymfactor-asymmetric. In the following we show that both our algorithms support a quality-runtime trade-off with respect to $\beta$. 
\subsection{Relaxed Generalized Christofides Algorithm}
For a given \asymfactor~we modify the algorithm presented in Section~\ref{sec:ezlk} by treating every \asymfactor-symmetric edge as symmetric. This formally results in a parameterization by the vertex cover of the subgraph induced by all $\beta$-asymmetric edges. We denote this parameter by $\ezlkparam_\beta$.
Since the \asymfactor-symmetric subgraph is not completely symmetric, the Christofides algorithm cannot be directly applied to it. Mart{\'{\i}}nez Mori and Samaranayake~\cite{bd-asymmetry} showed that the Christofides algorithm is $\beta\frac32$-stable by replacing every directed edge pair with an undirected edge before assigning it the cost of the more expensive direction. Combined with the arguments used for Theorem~\ref{thm:1.5kernel} this gives a $\frac{3}{2} \asymfactor$-additive lossy kernelization with respect to $\ezlkparam_\beta$.\par
This result can be  improved by turning the \asymfactor-symmetric subgraph into a symmetric graph that assigns the cost of the cheaper direction. Although this transformation may not result in a metric graph, it suffices that the original graph is metric to prove that an application of the Christofides algorithm yields a good approximation. The basic trick is to estimate the cost always with respect to the optimum solution for the original graph. The improvement of choosing this non-metric transformation originates from picking the better of the possible transversal directions, exploiting the fact that each replaced edge is more expensive only in one direction, which reduces the effect of the asymmetry $\asymfactor$ by a factor of $\frac 12$.
\begin{theorem}\label{thm:ezlk_relaxed}
    Metric ATSP admits a linear $\frac{3}{4}(1+\asymfactor)$-additive lossy kernelization for parameter~$\ezlkparam_\asymfactor$, size of a vertex cover for the graph induced by all $\asymfactor$-asymmetric edges, for any $\beta\geq 1$.
\end{theorem}

\subsection{Relaxed Generalized Tree Doubling Algorithm}
For the generalized tree doubling algorithm, we define a \asymfactor-one-way edge as a one-way edge that is \asymfactor-asymmetric. This way, the parameter counts only the number of \asymfactor-one-way edges in the spanning arborescence. We denote this reduced parameter by~$\algoparam_\asymfactor$. The alteration for the algorithm is straightforward, instead of deleting all one-way edges from $A$, we only delete \asymfactor-one-way edges. This results in fewer components and a smaller graph $\metaM_\asymfactor$. All further steps are analogous. The drawback is a change to the cost analysis: so far, we considered the component trees $T_i$ to be symmetric, as each backward edge was at most as expensive as the forward edge. The components trees $T_i$ contain no \asymfactor-one-way edge.
In other words, for every edge $(u,v) \in E[T_i]$, the backward edge $(v,u)$ can be at most $\asymfactor$ times as expensive.
In the adjusted tree doubling algorithm to compute the cheap path of the components, we use every edge in $T_i$ and its corresponding backward edge at most once, which overall results in a cost of at most $(1 + \asymfactor)c^*(G)$ for the paths through the components. Combined with the cost of at most $c^*(G)$ for an optimum tour through $\metaM_\asymfactor$, this yields the following.
\begin{theorem}
Metric ATSP can be $(2 + \asymfactor)$-approximated in $\mathcal{O}(2^{\algoparam_\asymfactor} {\algoparam}_\asymfactor^2 + n^2)$, where  $\algoparam_\asymfactor$ is the number of \asymfactor-one-way edges in a given minimum spanning arborescence, for any $\beta\geq 1$.
\end{theorem}

\section{Experimental Results}\label{sec:experiments}
In order to test the practical viability of the proposed algorithms, we implemented the them in their relaxed form as discussed in Section~\ref{sec:beta}. In this way, we observed how the algorithms behave when certain asymmetries are ignored. Further, in the implementation of the generalized Christofides algorithm we chose to solve the kernel exactly, as this lends to a better comparison with the generalized tree doubling algorithm. We evaluated and compared the performance on 19 asymmetric graph instances taken from the TSPLIB collection~\cite{reinelt_tsplibtraveling_1991}, the standard benchmark for TSP solvers.
\subsection{Implementation Details}
Our implementation is written in \emph{Python 3}, except for the vertex cover solver which is written in \emph{Java}. We used the Python library \emph{NetworkX}~\cite{hagberg_exploring_2008} for graph manipulation, the C++ library \emph{Lemon}~\cite{dezso_lemon_2011} for computing minimum spanning arborescences, and \emph{Concorde}~\cite{applegate_solution_1998} for solving TSP instances exactly. Since \emph{Concorde} is a TSP solver, we transformed the ATSP instances into a TSP instances with the transformation presented by Jonker and Volgenant~\cite{jonker_transforming_1983}. Our implementation is available on GitHub\footnote{\url{https://github.com/Blaidd-Drwg/atsp-approximation} (code will be uploaded for the final version)}.
\begin{table}[t]
    \centering
    \begin{tabular}{ccccc}
        \toprule
        instance name & \thead{symmetric \\ edges} & \thead{median \\ asymmetry factor} & \thead{maximum \\ asymmetry factor} & zero-cost edges \\
        \midrule
        br17 & 100\% & None & None & 12\% \\
        ft53 & 0\% & 2.04 & 23.04 & None \\
        ft70 & 0\% & 1.40 & 5.87 & None \\
        ftv170 & 6\% & 1.22 & 34.00 & None \\
        ftv33 & 6\% & 1.31 & 18.75 & None \\
        ftv35 & 5\% & 1.31 & 18.75 & None \\
        ftv38 & 6\% & 1.30 & 18.75 & None \\
        ftv44 & 5\% & 1.28 & 18.75 & None \\
        ftv47 & 3\% & 1.31 & 11.17 & None \\
        ftv55 & 5\% & 1.28 & 18.75 & None \\
        ftv64 & 4\% & 1.29 & 34.00 & None \\
        ftv70 & 4\% & 1.29 & 34.00 & None \\
        kro124p & 0\% & 1.04 & 3.42 & None \\
        p43 & 63\% & 13.61 & 14.64 & 3\% \\
        rbg323 & 33\% & 3.00 & 20.00 & 47\% \\
        rbg358 & 50\% & 3.00 & 18.00 & 65\% \\
        rbg403 & 49\% & 2.50 & 12.00 & 68\% \\
        rbg443 & 49\% & 2.67 & 11.00 & 69\% \\
        ry48p & 1\% & 1.04 & 3.63 & None \\
        \bottomrule
    \end{tabular}
    \caption{An overview of the 19 asymmetric TSPLIB instances and the properties of their metric closures. The asymmetry factor of each vertex pair $u, v$ is calculated by the formula $\max(\frac{c(u, v)}{c(v, u)}, \frac{c(v, u)}{c(u, v)})$ if there is no edge with cost zero between the two vertices and undefined otherwise. The median and maximum asymmetry factors were calculated by ignoring undefined values.}
    \label{tab:tsplib}
\end{table}
We note that the runtime of our implementations is incomparable to state of the art ATSP solvers. Among others, the reason is Python's inherently low performance and the inefficiency of solving ATSP with Concorde. However, this is of no importance for our evaluation of approximation ratio and parameter size and the proof of concept.

\subsection{Experiments}
Many instances in the TSPLIB are not metric, thus our algorithms cannot be directly applied. We therefore computed the metric closure of each graph by setting the cost of each edge $(u, v)$ to the cost of the shortest path from $u$ to $v$ and used this as input for our algorithm. The TSPLIB contains 19 asymmetric instances ranging from 17 up to 443 vertices with different underlying properties. The instances' names contain the number of vertices (e.g.\ \emph{ftv33}) and similar names indicate similar properties. For example, all instances starting with \emph{rbg} have relatively high symmetry and a high number of zero-cost edges, which distinguishes them from the other instances. Table~\ref{tab:tsplib} depicts an overview of the instances and the characteristics of their metric closures. In particular, the metric closure of \emph{br17} is completely symmetric, so we ignored it in our experiments.

\bgroup
\aboverulesep=0ex
\belowrulesep=0ex
\renewcommand{\arraystretch}{1.1}
\setlength\tabcolsep{0.64mm}
\begin{table}[tb]
\centering
\resizebox{\linewidth}{!}{\begin{tabu}
{@{}c|ccccc|[1pt]ccccc@{}}\toprule
& \multicolumn{5}{c|[1pt]}{\ezlk} & \multicolumn{5}{c}{\avtwo} \\
&100\%&25\%&6.25\%&1.56\%& 0\%&100\%&25\%&6.25\%&1.56\%& {0\%}\\ \midrule
{ft53} & 53/\textbf{1.00} & 29/1.54 & 13/1.70 & 6/1.69 & \textbf{1.72} & \textbf{45}/1.08 & \textbf{25}/\textbf{1.36} & \textbf{6}/\textbf{1.42} & \textbf{1}/\textbf{1.57} & 1.97
 \\
{ft70} & 69/1.02 & 34/1.24 & 12/1.26 & 7/1.41 & \textbf{1.24} & \textbf{64}/1.02 & \textbf{27}/\textbf{1.13} & \textbf{4}/\textbf{1.20} & \textbf{2}/\textbf{1.21} & 1.28
 \\
{ftv170} & 155/1.17 & 123/1.38 & \textbf{97}/1.57 & \textbf{64}/1.85 & 2.37 & \textbf{108}/\textbf{1.14} & \textbf{107}/\textbf{1.14} & 103/\textbf{1.21} & 75/\textbf{1.46} & \textbf{1.81}
 \\
{ftv33} & 29/\textbf{1.12} & 19/1.45 & 11/\textbf{1.43} & 5/1.56 & \textbf{1.33} & \textbf{19}/1.34 & \textbf{16}/\textbf{1.34} & 11/1.44 & \textbf{2}/\textbf{1.23} & 1.50
 \\
{ftv35} & 32/\textbf{1.07} & 21/1.51 & 12/1.55 & 6/1.49 & \textbf{1.38} & \textbf{23}/1.15 & \textbf{17}/\textbf{1.23} & \textbf{11}/\textbf{1.47} & \textbf{2}/\textbf{1.28} & 1.58
 \\
{ftv38} & 33/\textbf{1.13} & 23/1.38 & 12/\textbf{1.43} & 7/1.48 & \textbf{1.39} & \textbf{23}/1.24 & \textbf{18}/\textbf{1.33} & 12/1.54 & \textbf{3}/\textbf{1.30} & 1.62
 \\
{ftv44} & 40/\textbf{1.11} & 32/1.41 & 19/1.46 & 10/1.56 & \textbf{1.54} & \textbf{32}/1.24 & \textbf{25}/1.41 & \textbf{18}/\textbf{1.41} & \textbf{7}/\textbf{1.50} & 1.79
 \\
{ftv47} & 44/\textbf{1.05} & 32/1.47 & 19/1.66 & 13/1.65 & 1.66 & \textbf{35}/1.09 & \textbf{30}/\textbf{1.16} & 19/\textbf{1.34} & \textbf{9}/\textbf{1.38} & \textbf{1.58}
 \\
{ftv55} & 49/\textbf{1.13} & 38/1.44 & \textbf{23}/1.57 & 15/1.65 & \textbf{1.84} & \textbf{37}/1.20 & \textbf{32}/\textbf{1.26} & 25/\textbf{1.34} & \textbf{12}/\textbf{1.58} & 2.00
 \\
{ftv64} & 57/1.11 & 46/1.46 & \textbf{30}/1.66 & 18/1.73 & 1.72 & \textbf{50}/\textbf{1.10} & \textbf{43}/\textbf{1.15} & 31/\textbf{1.29} & \textbf{14}/\textbf{1.71} & \textbf{1.45}
 \\
{ftv70} & 63/\textbf{1.11} & 50/1.43 & \textbf{32}/1.64 & 20/1.72 & 1.96 & \textbf{53}/1.26 & \textbf{47}/\textbf{1.14} & 33/\textbf{1.22} & \textbf{16}/\textbf{1.57} & \textbf{1.51}
 \\
{kro124p} & 99/1.11 & 86/1.30 & 65/1.36 & 40/1.41 & \textbf{1.24} & \textbf{81}/\textbf{1.06} & \textbf{70}/\textbf{1.13} & \textbf{57}/\textbf{1.20} & \textbf{34}/\textbf{1.28} & 1.37
 \\
{p43} & 15/1.01 & 6/1.01 & 0/1.01 & 0/1.01 & 1.01 & \textbf{0}/1.01 & \textbf{0}/1.01 & 0/1.01 & 0/1.01 & 1.01
 \\
{rbg323} & \textbf{148}/\textbf{1.02} & 59/\textbf{1.17} & 43/\textbf{1.19} & 18/1.30 & 1.34 & 235/1.09 & \textbf{22}/1.27 & \textbf{6}/1.27 & \textbf{0}/1.30 & \textbf{1.30}
 \\
{rbg358} & \textbf{108}/\textbf{1.01} & 47/\textbf{1.13} & 27/\textbf{1.15} & 22/\textbf{1.14} & \textbf{1.18} & 232/1.03 & \textbf{39}/1.14 & \textbf{18}/1.19 & \textbf{13}/1.20 & 1.22
 \\
{rbg403} & 125/\textbf{1.01} & 41/\textbf{1.12} & 11/1.26 & 11/1.26 & \textbf{1.17} & \textbf{113}/1.05 & \textbf{30}/1.14 & \textbf{0}/\textbf{1.24} & \textbf{0}/\textbf{1.24} & 1.24
 \\
{rbg443} & 138/\textbf{1.00} & 43/\textbf{1.14} & 12/1.24 & 12/1.24 & \textbf{1.15} & \textbf{127}/1.04 & \textbf{32}/1.17 & \textbf{0}/1.24 & \textbf{0}/1.24 & 1.24
 \\
{ry48p} & 47/1.20 & 37/1.40 & 23/1.46 & 11/1.47 & \textbf{1.16} & \textbf{28}/\textbf{1.10} & \textbf{22}/\textbf{1.14} & \textbf{11}/\textbf{1.24} & \textbf{5}/\textbf{1.29} & 1.21
 \\
\bottomrule
\end{tabu}}
\caption{Summary of experimental results. Rows represent the TSPLIB instances, columns represent experiments with the percentage of asymmetric edges that were treated as asymmetric shown in the column header.
Each cell contains kernel size and approximation factor, separated by a slash.
In the 0\% column the kernel size was omitted since it is always 0.
Values indicating superiority are set in bold font.}
\label{tab:results}
\end{table}
\egroup

For each instance we executed each algorithm five times with different values for $\asymfactor$ starting with $\asymfactor$ set to 1, which corresponds to 100\% of the asymmetric edges to be considered asymmetric. In the following 3 experiments the value of $\asymfactor$ was raised each time, reducing the number of edges considered asymmetric to a quarter of the previous experiment.
Some instances include many zero-cost edges, so there is no value of $\asymfactor$ ignoring those. We hence considered zero-cost edges to have a small positive cost (set to 0.1) when calculating the asymmetry factor for relaxation. This way, we implemented a relaxation that also ignores edges with a small additive error in case of these otherwise undauntedly asymmetric one-way edges of cost~0. Note that we did not alter the instance, but only used these additive errors for relaxation decisions.
Finally, $\asymfactor$ was set to $\infty$, such that the graph is treated as symmetric. This is equivalent to running the non-generalized versions of the tree doubling  and the Christofides algorithm.

For each of the experiments we evaluated the approximation factor, i.e.\ the cost of the tour compared to the optimum. Additionally, we evaluated the size of the graph on which we solve ATSP exactly, i.e.\ the size of the parameter for each algorithm. For simplicity we refer to both of these values as \emph{kernel size}. The results are shown in Table~\ref{tab:results}. For more detailed plots of the results, refer to Appendix~\ref{app:results}.

\subsection{Evaluation}
First of all we note that most graphs contain very little symmetry.
This leads to large kernel sizes for $\asymfactor=1$, i.e.\ only some graphs with more than $10\%$ symmetry have kernel sizes below $50\%$ of the original graph size.
Still, we observe that the approximation factor is always far below the upper bound, never exceeding even $2.0$.
Also, we observe that interpolating $\asymfactor$ to reduce the number of relevant asymmetric edges produces a valuable trade-off between approximation quality and kernel size.
Comparing both algorithms, we observe that on the majority of instances and values for $\asymfactor$ the generalized tree doubling algorithm produces smaller kernels.
\par
These results underline the practicality of our approach, especially with regards to the kernel sizes obtained by choosing a suitable $\asymfactor$ as well as the achieved approximation ratios.
Still, the significance of the results is very limited due to the small dataset size and questionable representativeness of the instances.
An evaluation of the algorithms on bigger datasets and representative real-world instances remains open.

\bibliographystyle{plain}
\bibliography{tsp}

\newpage
\appendix
\section{Omitted Proofs}\label{app:lemmata}
\paragraph*{Proof of Lemma~\ref{lemma:induced_graph_metric}}
    Let $G$ be a metric graph and $V' \subseteq V[G]$.
    Then, $G[V']$ is metric as well.
    \begin{proof}
        Proof by contradiction.
        Suppose that $G[V']$ is not metric.
        Then there exist three vertices $u, v, w \in V'$ for which $c(u,v) + c(v,w) < c(u, w)$.
        Since every vertex and edge of an induced subgraph exists in the original graph, the triangle $u, v, w$ violates the triangle inequality in $G$.
        This contradicts our original assumption that $G$ is metric.
    \end{proof}

\paragraph*{Proof of Lemma~\ref{lemma:induced_graph_has_leq_tour}}
 Let $G$ be a metric ATSP instance and $V' \subseteq V[G]$. Then, $c^*(G[V']) \leq c^*(G)$.

    \begin{proof}
 Starting with an optimum solution $\tau$ for $G$, iteratively remove vertices $v \notin V'$ from $\tau$ until it contains only vertices from $V'$.
        Removing a vertex is equal to performing a metric shortcut and thus does not increase the cost of the whole tour.
        The claim follows.
    \end{proof}

\paragraph*{Proof of Theorem \ref{thm:1.5kernel}}
    Metric ATSP admits a linear $1.5$-additive lossy kernelization for parameter $\ezlkparam$, size of a vertex cover of the subgraph induced by all asymmetric edges.
    \begin{proof}
        The kernelization algorithm reduces a metric ATSP instance $(G, \ezlkparam)$ with a given vertex cover $\mathit{VC}$ for $G[E_a]$ of size $\ezlkparam$ to an instance $(G', \ezlkparam)$. It chooses arbitrarily a vertex $v\in V[G]\setminus \mathit{VC}$ and sets $G'=G[\mathit{VC}\cup \{v\}]$ which is a linear reduction for parameter $\ezlkparam$.\par
    The solution lifting algorithm takes a metric ATSP instance $G$, the corresponding kernel $G'$ and a $\gamma$-approximate TSP tour $\tau'$ for $G'$. It returns a TSP tour~$\tau$ for $G$.
The algorithm can be described as follows.
\begin{enumerate}
    \item Choose the same vertex $v \in V[G'] \setminus V_a[G]$ as chosen by the kernelization algorithm.
    \item Compute a 1.5-approximate tour $\tau''$ for  $G''=G\left[V[G]\setminus (V[G'] \setminus \{v\})\right]$ with the Christofides algorithm.
    \item Construct and return $\tau$ by inserting $\tau'$ into $\tau''$ at $v$ and taking metric shortcuts.
\end{enumerate}

In step 2, the Christofides algorithm is used to compute an approximate tour on a subgraph.
This is possible because the vertices in the subgraph $G''$ are the complement of a vertex cover for $G[E_a]$, which means that all edges among these vertices are symmetric.
Hence, the Christofides algorithm receives a symmetric subgraph as input.
This yields a $\frac{3}{2}$-approximation $\tau''$ for $G''$, and by Lemma~\ref{lemma:induced_graph_has_leq_tour} we obtain $c(\tau'') \leq \frac{3}{2} \cdot c^*(G)$.

In step 3 we make use of the fact that $v$ appears both in $\tau'$ and $\tau''$ to combine the two tours into a single circuit.
This circuit uses every edge in $\tau'$ and $\tau''$ exactly once.
Therefore, the cost of the circuit is $c(\tau') + c(\tau'') \le \gamma c^*(G) + \frac{3}{2}c^*(G)$.
We then take metric shortcuts to turn the circuit into a cycle $\tau$.

The cycle $\tau$ visits every vertex visited by $\tau'$ and $\tau''$.
Since $\tau'$ and $\tau''$ span all vertices of $G$ $\tau$ is a tour of $G$ with cost at most
\begin{equation}
\label{eq:clk_cost}
c(\tau) \le \left(\gamma + \frac{3}{2} \right) \cdot c^*(G).
\end{equation}
 Hence the solution lifting algorithm turns a $\gamma$-approximate solution for $G'$ into a $(\gamma+1.5)$-approximate solution for $G$.
     At last, both algorithms run in polynomial time.
    \end{proof}

\paragraph*{Proof of Lemma~\ref{lemma:tspm_leq_tspg}}
    Let $G$ be a metric ATSP instance and $\metaM$ be a minor of $G$. Then $c^*(\metaM) \le c^*(G)$.
    \begin{proof}
        Consider an arbitrary sequence of contractions of $G$ that result in $\metaM$.
        Then, we map each vertex $v \in \metaM$ to the subset of vertices $S_v \subseteq V[G]$ that $v$ was contracted from. Let $\tau^*$ be an optimal tour for $G$.
        We remove vertices from $\tau^*$ until it contains exactly one vertex from each vertex set $S_v$.
        Let $\tau$ be the remaining tour.
        Since we only removed vertices and $G$ is metric, $\tau$ is at most as expensive as $\tau^*$.

        Let us consider an arbitrary edge $(x,y)$ in $\tau$ and let $x$ be in $S_v$ and $y$ in $S_w$.
        It follows by the construction of $\metaM$ that $\metaM$ contains an edge $(a, b)$ with $a \in S_v,\, b \in S_w$, for which $c(a, b) \leq c(x, y)$.
        The respective edge in $\metaM$ for every edge in $\tau$ yields a tour in $\metaM$ whose cost is at most the cost of $\tau$, and therefore at most $c(\tau^*) =c^*(G)$.
    \end{proof}

\setcounter{algocf}{0}
\begin{algorithm}
\caption{\avtwo}
    \DontPrintSemicolon
    $T_1,\dots,T_\compcount \gets$ components of $A[\{e \in E[A] \mid e$ is not a one-way edge$ \}]$\;
    $\metaM \gets$ complete graph with vertices $\metam_1, \dots, \metam_\compcount$\;
    \ForEach{$\metam_i, \metam_j \in V[M]$ with $i \neq j$}{
        $c(\metam_i, \metam_j) \gets \min\left(\left\{ c(t_i, t_j) \mid t_i \in V[T_i], t_j \in V[T_j] \right\}\right)$\;
    }
    $\tau' \gets$ an optimum tour of $M$\;
    \ForEach{subsequent $\metam_{i-1}, \metam_i, \metam_{i+1} \in \tau'$}{
        $v^{T_{i-1}}_{out}, v^{T_i}_{in} \gets \mathrm{argmin}_{t_{i-1} \in V[T_{i-1}], t_i \in V[T_i]}(c(t_{i-1}, t_i))$\;
        $v^{T_i}_{out}, v^{T_{i+1}}_{in} \gets \mathrm{argmin}_{t_i \in V[T_i], t_{i+1} \in V[T_{i+1}]}(c(t_i, t_{i+1}))$\;
        $\chi_i \gets $ path from $v^{T_i}_{in}$ to $v^{T_i}_{out}$ spanning $T_i$  by adjusted tree doubling\;
    }
    $\tau \gets$ concatenation of $(\chi_1,\dots,\chi_\compcount)$\;
    \Return $\tau$
\end{algorithm}

\paragraph*{Proof of Theorem~\ref{thm:3_approx}}
    A 3-approximation for metric ATSP can be computed in time $\mathcal{O}(2^\algoparam \cdot \algoparam^2 + n^2)$, where $n$ is the number of vertices in the input graph $G$ and $\algoparam$ is the number of one-way edges in a given minimum spanning arborescence.
    \begin{proof}
For a metric ATSP instance $G$ and a minimum spanning arborescence $A$ for $G$, we claim that Algorithm~\ref{algo_v2_main} run on $G$ with $A$ yields a tour $\tau = (\chi_1, \dots, \chi_{\compcount})$ of cost at most 3 times the optimum. Since $\chi_i$ spans the vertices in $T_i$, $1\leq i \leq \compcount$ and $\tau$ is built from a concatenation of all these paths, $\tau$ visits each vertex in $\bigcup_{i=1}^\compcount V(T_i)=V(G)$ exactly once and is thus a tour of $G$.
        In total, $\tau$ consists of the paths through each component in addition to the edges $(v_i^{out}, v_{i+1}^{in})$ for $i \in [1, \compcount]$, which correspond to the edges in an optimal solution for ${\metaM}$.
        Since the subtours $\chi_i$ through the components $T_i$ are computed by the adapted tree doubling algorithm, they satisfy the following equation      
    \begin{equation}\label{eq:sum_chi_leq_2tsp}
        \sum_{i=1}^{\compcount} c(\chi_i) \leq  \sum_{i=1}^{\compcount} 2 \cdot c(T_i) \leq 2c^*(G) .
    \end{equation} 
        Together with Lemma~\ref{lemma:tspm_leq_tspg}  it follows that the constructed tour is at most 3 times as expensive as the optimum tour for ${G}$.

        The runtime of Algorithm~\ref{algo_v2_main} is dominated by step~\ref{algo:line:solve_M} which is solving an ATSP instance of size $\compcount$ and thus has a runtime exponential in $\algoparam$, unless $P=\mathit{NP}$.
        In particular, a solution that uses the dynamic programming algorithm by Held and Karp~\cite{held_dynamic_1962} runs in $\mathcal{O}(2^{\algoparam} {\algoparam}^2)$.
        The runtime of all remaining steps is polynomial in $n$, where the most expensive operation is the contraction of $G$ into the graph $\metaM$, which has quadratic runtime.
        The overall runtime of the algorithm is therefore $\mathcal{O}(n^2 + 2^{\algoparam} \algoparam^2) \subseteq \mathcal{O}^*(2^\algoparam)$.
    \end{proof}

\paragraph*{Proof of Proposition~\ref{prop:always_minor_of_metric_short}}

    Let $G$ be a complete, directed graph with cost function $c$.
    Then, there exists a complete, metric graph $\hat{G}$ of which $G$ is a minor.
    \begin{proof}
Given a complete, directed graph $G$ with cost function $c$, we construct a metric graph $\hat G$ as follows.
        For each vertex pair $\{ u, v \} \in V[G] \times V[G]$ with $u \neq v$, we create  vertices $u_v$ and $v_u$ in $\hat{G}$.
        We create  edges $(u_v, v_u)$ and $(v_u, u_v)$, and assign them the costs $c(u, v)$ and $c(v, u)$, respectively.
        Finally, we connect every vertex pair that is not yet connected with an edge with cost $m+1$, where $m$ is the maximum edge cost in $G$.
        Every vertex in $\hat{G}$ is incident to exactly one edge whose cost is not $m+1$.
        Therefore, every triangle in $\hat{G}$ consists either of three edges with cost $m+1$, or of two edges with cost $m+1$ and one edge with cost less than $m+1$.

        Every vertex $u$ in $G$ is represented by a set of vertices $V_u$ in $\hat{G}$: $V_u = \{ u_v \mid v \in V[G], v \neq u \}$.
        Similar to a vertex $u$ having exactly one edge to every vertex $v$ in $G$, a vertex set $V_u$ has exactly one edge with cost $c(u,v) < m+1$ to every vertex set $V_v$ in $\hat{G}$, while all other edges that begin in $V_u$ have cost $m+1$.
        In particular, for every edge $(u,v)$ in $G$ there is exactly one edge from $V_u$ to $V_v$ in $\hat{G}$ with the same cost.

        In order to transform $\hat{G}$ into $G$ we iteratively contract all the vertices in each set $V_u$.
        We show that during these contractions an edge with cost less than $m+1$ is never deleted.
        When contracting two vertices $x, y \in V_u$, an edge can be deleted for two reasons:
        \begin{enumerate}
            \item The edge connects $x$ and $y$.
            In this case, the edge connects vertices from the same set $V_u$ and has cost $m+1$.
            \item The edge connects $x$ and some vertex $w$, w.l.o.g.~in the direction from $x$ to $w$, and $c(y, w) \le c(x, w)$.
            The vertices $x$ and $y$ belong to the same vertex set, thus there is at most one edge (in each direction) with cost less than $m+1$.
            Since $c(y, w) \le c(x, w)$, the edge $(x, w)$ must have cost $m+1$.
            The case of an edge from $w$ to $x$ is analogous.
        \end{enumerate}

        After contracting all $n$ vertex sets $V_i$, there are exactly $n$ vertices (and thus $n^2 - n$ edges) left.
        Since we did not delete any edge with cost less than $m+1$ and $G$ has $n^2 - n$ edges as well, we know that there cannot be any edge with cost $m+1$ left in the contracted graph.

        In the following we show that the contractions of $\hat{G}$ result in $G$.
        Consider an arbitrary edge $(u,v) \in E[G]$.
        Let $c_u$ be the vertex obtained by contracting all vertices in $V_u$.
        We know that all edges from $V_u$ to $V_v$ have cost $m+1$, except exactly one edge with cost $c(u,v)$.
        Since contractions always keep the cheaper edge, the cost of the remaining edge $(c_u, c_v)$ also has cost $c(u, v)$.
        As a result, each $c_u$ in the contracted graph can be considered as $u$ in $G$.
    \end{proof}

\paragraph*{Proof of Theorem~\ref{thm:3kernel}} 
    Metric ATSP admits a linear $2$-additive lossy kernelization with respect to  parameter \algoparam, number of one-way edges in a minimum spanning arborescence.
    \begin{proof}
Consider a parameterized metric ATSP instance $(G, \algoparam)$ with minimum spanning arborescence $T$. Let $T_1,\dots , T_{\algoparam +1}$ be the connected components of the graph created from removing all one-way edges from $T$.
Pick arbitrarily one vertex $v^t_i$ from $T_i$ for each $i\in \{1,\dots, \algoparam+1\}$ and consider as kernel $G'=[\{v_1^t, \dots , v_{k+1}^t\}]$, the subgraph induced by these vertices.
Since $G'$ has only $\algoparam+1$ vertices, any spanning tree on $G'$ has $\algoparam$ edges and thus at most $\algoparam$ one-way edges in total, so $\algoparam'\leq \algoparam$. Further, as induced subgraph of $G$, $G'$ is a metric instance of ATSP (in contrast to the graph $\metaM$ used for Theorem~\ref{thm:3_approx}) which shows that $(G',\algoparam')$ is a linear kernel with respect to $\algoparam$. \par
        The solution lifting algorithm turns a $\gamma$-approximate solution for $(G',\algoparam')$ into a solution for $(G,\algoparam)$ by replacing each vertex $v_i^t$ by a tour through the vertices in $T_i$. These tours are computed by tree doubling for $T_i$. Since $G'$ is a subgraph of $G$, an optimal tour for $G'$ is at most as expensive as an optimal tour for $G$. The cost of the lifted solution is hence at most $\gamma c^*(G')\leq \gamma c^*(G)$ for the tour connecting the components, plus $2c^*(G)$ for the subtours for each $T_i$, which overall is a $(\gamma+2)$-approximate solution.
        Both algorithms run in polynomial time.
    \end{proof}

\paragraph*{Proof of Theorem~\ref{thm:ezlk_relaxed}}
    Metric ATSP admits a linear $\frac{3}{4}(1+\asymfactor)$-additive lossy kernelization for parameter~$k_\asymfactor$, the cardinality of a vertex cover for the graph induced by the $\asymfactor$-asymmetric edges.
    \begin{proof}
Let $G$ be a \asymfactor-symmetric metric graph and let $G_<$ be the undirected graph created from $G$ by replacing each pair of directed edges $(u, v), (v, u)$ with cost $c_1$ and $c_2$ in $E[G]$ by an undirected edge between $u$ and $v$ with cost $\min\{c_1,c_2\}$. Let $T$ be a minimum spanning tree for $G$. Since all edge costs in $G_<$ are smaller or equal to the costs in $G$, the cost of $T$ is at most the cost of a minimum spanning tree for $G$ and hence at most $c^*(G)$.\par
 Let $V'\subseteq V[G_<]$ be the set of vertices with odd degree in $T$. Since the original graph $G$ is metric, a min-cost perfect matching for $V'$ in $G$ has a cost of at most $\frac12c^*(G)$. Again, the cost of such a matching in $G_<$ is smaller or equal to its corresponding cost in $G$, hence a min-cost perfect matching $M$ of $V'$ in $G_<$ has cost at most $\frac12c^*(G)$. Let $W$ be an Eulerian circuit for the edges in $T\cup M$. Consider both directions to traverse $W$ in $G$ and the resulting cost of the tour. Each edge in $W$ that is directed in $G$ has a larger cost than the undirected edge in $G_<$ only in one of the traversal directions.  In the worst case, all edges in $W$ correspond to directed edges in $G$ and have in one direction a cost of $\asymfactor$ times the cost of their counterparts in $G_<$. The better of the two traversal directions hence corresponds to a tour of cost at most $\frac12(1+\asymfactor)$ times the cost of $W$ in $G_<$, which is at most $C(T\cup M)\leq \frac{3}{2} \cdot c^*(G)$. In the metric graph $G$, metric shortcuts can be used to turn $W$ into a proper TSP tour without increasing the cost, which overall yields a $\frac{3}{4}(1+\asymfactor)$-approximate solution.\par
Applying this  $\frac{3}{4}(1+\asymfactor)$-approximate solution for the  \asymfactor-symmetric subgraph of a given ATSP instance as solution lifting algorithm in Theorem~\ref{thm:1.5kernel} yields the claimed lossy kernelization.
    \end{proof}

\section{Tightness for ratio 2.5 in Corollary~\texorpdfstring{\ref{cor:2.5apx}}{ 3.2}}\label{sec:2.5_tightness}
\begin{figure}
    \centering
    \includegraphics[page=4,width=0.5\textwidth,trim={4cm, 6cm, 4cm, 5.5cm},clip]{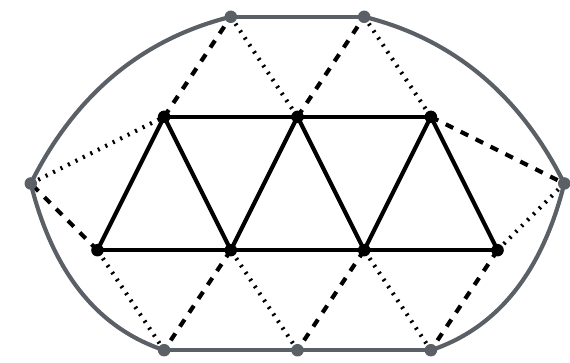}
    \caption{The constructed graph $G_k$ for $k=7$.
    Black and gray edges are symmetric with cost 2.
    Dotted edges are symmetric with cost 1.
    Dashed edges are asymmetric, with cost 1 from gray to black vertex and cost 2 from black to gray vertex.}
    \label{fig:2.5tight_figure}
\end{figure}

In the following we show the tightness of the ratio of 2.5 for the approximation for ATSP given in Corollary~\ref{cor:2.5apx}. Recall that on input $G$, this algorithm performs the following steps:
\begin{enumerate} 
\item Compute a minimum vertex cover $VC$ for the subgraph induced by all asymmetric edges of $G$.
\item Pick an arbitrary vertex $v\in V[G]\setminus VC$ and compute a minimum tour $\tau'$ for the subgraph $G[VC\cup\{v\}]$.
\item Compute a tour $\tau''$ for $G[V[G]\setminus VC]$ with Christofides algorithm.
\item Append $\tau'$ and $\tau''$ at $v$ into a tour for $G$.
\end{enumerate}
We construct a family of graphs $(G_k)_k$ for which. in the worst case, our algorithm finds a tour of cost 2.5 times the optimum as $k$ approaches infinity. The idea is the following: The graph $G_k$ consists of two symmetric cycles, a \emph{gray} and a \emph{black} cycle, of length $k$, which are connected by cheap asymmetric edges.
An optimum tour uses the asymmetric edges and contains no edge from the cycles; however by choosing the gray cycle as vertex cover in step 2, we ensure that none of these edges can be used by our algorithm.
Furthermore, the black cycle contains additional edges which turn it into a worst case instance for the Christofides algorithm used in step 3.
In  Figure~\ref{fig:2.5tight_figure}, $G_7$ is shown as an example.

\subsection{First Step: Creating an Incomplete Graph}\label{sec:2.5_tightness_incomplete_graph}
We now describe the exact construction of the graph $G_k$.
It consists of $k$ \emph{gray} vertices $g_1, \dots, g_k$ and $k$ \emph{black} vertices $b_1, \dots, b_k$.
The gray vertices form a cycle $(g_1, \dots, g_k)$, which we denote by \emph{gray cycle}.
The black vertices form a second cycle $(b_1, \dots, b_k)$, which we denote by \emph{black cycle}.
All edges in the gray and black cycles are symmetric and have cost 2.
Furthermore, the black cycle has a series of additional internal edges.
These form the path $b_1, b_k, b_2, b_{k-1}, \dots, b_{\lceil k/2 \rceil}$, which alternates between vertices at the start and at the end of the cycle, and ends in a vertex in the middle.
All edges on this path are symmetric and have cost 2.
Note that this path thus forms a minimum spanning tree of the subgraph induced by the black vertices, and that the start and end vertices of the path are as distant as possible: this constitutes a worst case instance for the Christofides algorithm.

In order to allow an optimum tour to alternate between the gray and black cycle, the gray cycle is connected to the black cycle with a number of asymmetric edges: for $1 \le i \le k$ there is an edge $(b_i, r_i)$ with cost 2 and an opposite edge with cost 1.
Additionally, $b_i$ is connected to $r_{i+1}$ by a symmetric edge of cost 1 for $1\le i \le k$.

The cycle $g_1, b_1, \dots, g_k, b_k$, which alternates between the gray and black vertices, is in fact a tour.
The cost of this cycle is $2k$, as it contains only edges with cost 1 and consists of $2k$ vertices.
Since there are no edges with cost less than 1, the tour is an optimal solution.

We note that the set of gray vertices is a minimum vertex cover.

\subsection{Making the Graph Complete}\label{sec:2.5_tightness_complete_graph}
The graphs $G_k$ are not complete, and thus not a valid metric ATSP instance.
In order to apply our algorithm, we must first insert all missing edges and ensure that the triangle inequality is not violated.
First, we define a generalized form of the triangle inequality for incomplete graphs, which we call \emph{polygon inequality}.
Then, we show that every graph $G_k$ satisfies the polygon inequality. Finally, we show that every graph which satisfies the polygon inequality can be transformed into a complete graph which satisfies the triangle inequality without modifying the existing edge costs.

\begin{definition}
    In a graph $G$, the polygon inequality holds if and only if any path $P$ between two vertices $u$ and $v$ is at least as expensive as the direct edge $(u,v)$:
    \begin{equation}\label{eq:polygon_inequality}
        \forall v_1, \dots, v_j \in V[G]^*: \sum\limits^{j-1}_{i=1} c(v_i, v_{i+1}) \ge c(v_1, v_j)
    \end{equation}
\end{definition}

Note that in complete graphs, the polygon inequality and the triangle inequality are equivalent.

We now show that the polygon inequality holds in $G_k$ and that any strongly connected graph, where the polygon inequality holds, can be turned into a complete graph, where the triangle inequality holds as well.

\begin{proposition}\label{prop:polygon_inequality_holds_in_G_k}
    Let $G_k$ be a graph constructed as described in Section~\ref{sec:2.5_tightness_incomplete_graph}.
    Then, eq.~\ref{eq:polygon_inequality} holds in $G_k$.
    \begin{proof}
        Let $P_{uv}$ be an arbitrary path from $u$ to $v$ with $u,v \in V[G_k]$.
        There are two cases to consider:
        \begin{enumerate}
            \item $P_{uv}$ consists of a single edge.
                In this case the path consists only of the edge $(u,v)$.
                Thus, $C(P_{uv}) = c(u,v)$, which satisfies the polygon inequality.
            \item $P_{uv}$ consists of at least two edges.
                The cheapest edge in $G_k$ has cost 1, thus $C(P_{uv}) \ge 2$.
                The most expensive edge in $G_k$ has cost 2, so $2 \ge c(u,v)$.
                As a result, $C(P_{uv}) \ge c(u,v)$, which satisfies the polygon inequality.
        \end{enumerate}
    \end{proof}
\end{proposition}

\begin{lemma}\label{lemma:polygon_inequality_implies_triangle_inequality}
    Let $G$ be an incomplete, directed, and strongly connected graph in which the polygon inequality (eq.~\ref{eq:polygon_inequality}) holds.
    Then, $G$ can be turned into a complete, directed graph $G'$ in which the triangle inequality holds without modifying the existing edge costs.

    \begin{proof}
        Since the polygon inequality (eq.~\ref{eq:polygon_inequality}) implies the triangle inequality in complete graphs, it suffices to show that we can make the graph complete without violating the polygon inequality.
        To do so, missing edges are iteratively inserted and assigned the cost of the cheapest path between the connected vertices that existed in the graph prior to the insertion.
        We show that this never violates the polygon inequality by induction over the number of added edges.

        As the base case, we know from Proposition~\ref{prop:polygon_inequality_holds_in_G_k} that the polygon inequality holds in $G$.

        For the induction step, we choose two arbitrary vertices $u,v$ that are not yet connected by an edge $(u,v)$ and compute a cheapest path $P_{uv}$ from $u$ to $v$.
        Since $G$ is strongly connected, we can be sure that $P_{uv}$ always exists.
        We insert the edge into $G'$ with $c(u, v) = C(P_{uv})$ and show that inserting $(u, v)$ does not violate the polygon inequality.
        For the insertion to violate the inequality, one of two cases would have to apply:
        \begin{enumerate}
            \item
                There exists a path $u, \dots, v$ which has lower cost than $(u, v)$.

                This cannot occur, as the cost of $(u, v)$ was chosen equal to the cost of the cheapest path from $u$ to $v$.
            \item
                $(u, v)$ is part of a path $P_{xy}$ between two vertices $x$ and $y$ and $C(P_{xy}) < c(x, y)$.

                We show that this cannot occur as well. Already before the insertion of $(u, v)$ there existed a path $P'_{xy}$, which can be obtained from $P_{xy}$ by replacing $(u, v)$ with a cheapest path from $u$ to $v$ and taking a metric shortcut.
                By to the choice of $c(u, v)$ it follows that $C(P_{xy}) = C(P'_{xy})$.
                We know from the induction hypothesis that the polygon inequality holds for any path not containing $(u, v)$, and thus $C(P'_{xy}) \ge c(x,y)$.
                It follows that the inequality also holds for $P_{xy}$.
        \end{enumerate}
        In both cases, the polygon inequality still holds after inserting the new edge.
        The claim follows.
    \end{proof}
\end{lemma}

Note that every cheapest path in $G$ is still a cheapest path in $G'$. The reason is that for every new edge $(u,v)$, $G$ already contained a path $u, \dots, v$ with equal cost.

\subsection{Analysis of the Approximation Ratio}
Let $G_k$ be the incomplete graph constructed as described in Section~\ref{sec:2.5_tightness_incomplete_graph} for a given $k$, and $G'_k$ be its corresponding complete version as described in Section~\ref{sec:2.5_tightness_complete_graph}.
In the following we analyze the approximation given by  Corollary~\ref{cor:2.5apx} on the graph $G'_k$.

An optimum tour of $G'_k$ can be found by alternating between vertices of the black and gray cycle $b_1, g_1, \dots, b_k, g_k$.
This leads to a tour of length $2k$.
Since the cheapest edge in $G_k$ has cost 1 and there are $2k$ vertices, no tour can be cheaper than $2k$.

The first step of our algorithm is to find a minimum vertex cover on the graph induced by the asymmetric edges.
Suppose that the gray vertices are chosen as the minimum vertex cover.
The set of gray vertices indeed is a minimum vertex cover as can be seen as follows:
\begin{enumerate}
    \item The gray vertices form a vertex cover:
        This is equivalent to stating that the black vertices form an independent set, thus there are no asymmetric edges between any of the black vertices.
        For this, recall that all edges between black vertices are symmetric in $G_k$.
        Further, as can be seen in  Figure~\ref{fig:2.5tight_figure}, for every pair of black vertices $u, v$ there always is a cheapest path $P^*_{u,v} = u, \dots, v$ that uses only black vertices, and thus $c(P^*_{u,v})$ equals $c(P^*_{v,u})$.
        As a result, all edges between black vertices in $G'_k$ must be symmetric as well.
    \item The vertex cover formed by the gray vertices is minimal: There is an asymmetric edge between every corresponding pair of vertices $b_i$ and $r_i$.
        In order to cover those edges, at least one vertex per edge needs to be taken into the vertex cover, therefore there cannot be a vertex cover with fewer than $k$ vertices.
\end{enumerate}

Step 1 of the algorithm computes an optimal solution for the subgraph on one black and $k$ gray vertices.
It follows immediately  that there is an optimum solution for this subgraph which traverses almost the whole gray cycle once, only leaving once to pick up the black vertex.
The cost of this solution is $2k$.

Step 3 runs the Christofides algorithm on the subgraph on the black vertices. We assume an unlucky choice of the MST, namely the path we added to the black cycle in $G_k$.
The path has cost of $2k - 2$.
Since we need $k - 1$ edges for a spanning tree and each edge between black vertices has at least cost 2, we know that our path is an MST.
The MST has only two vertices with odd degree: start and end vertex of the path, which have the biggest possible distance to each other (which is $\lfloor \frac{k}{2} \rfloor$).
As a result, the edge chosen in the matching step has cost $2 \lfloor \frac{k}{2} \rfloor$, which is at least $k - 1$.
The total cost of the solution returned by the Christofides algorithm for step 3 is thus at least $2k - 2 + k - 1$.

Overall, this leads to a cost of at least $5k - 3$.
When compared to the optimum tour, we obtain a lower bound of $\frac{5k - 3}{2k}$ for the approximation ratio.
As $k$ approaches infinity, this ratio converges to 2.5.

\section{Adapted Tree Doubling Algorithm}\label{sec:algo_v2_sub}
\begin{algorithm}[htbp]
\caption{Adjusted tree doubling for a spanning path of component $T_i$}
\label{algo_v2_sub}
    \KwIn{Symmetric tree $T_i$, vertices $v_{in}^{T_i}, v_{out}^{T_i} \in T_i$}
    \KwOut{A spanning path}
    \DontPrintSemicolon
    $P_i \leftarrow$ the unique path from $v^{T_i}_{in}$ to $v^{T_i}_{out}$ in $T_i$\;
    $T^\metaM_i := T_i$\;
    \ForEach{edge $e$ in $T_i$} {
        \If{$e \notin P_i$} {
            double $e$ in $T^\metaM_i$\;
        }
    }
    $\pi_i \leftarrow$ an Eulerian trail from $v_{in}^{T_i}$ to $v_{out}^{T_i}$ in $T^\metaM_i$\;
    $\chi_i \leftarrow$ the metric shortcut of $\pi_i$\;
    return $\chi_i$\;
\end{algorithm}
Algorithm~\ref{algo_v2_sub} contains a more formal description of how the cheap paths through the components are computed.
Note that the algorithm assumes the component $T_i$ to be symmetric.
This is a pessimistic assumption, because the edges of the tree are at least as expensive in the direction used by the tree as in the opposite direction.
As this pessimistic assumption still yields paths that are at most twice as expensive as the entire component, it is valid to treat the components as symmetric.

\section{Experiment Plots}\label{app:results}
\begin{figure}
    \includegraphics[width=0.97\textwidth]{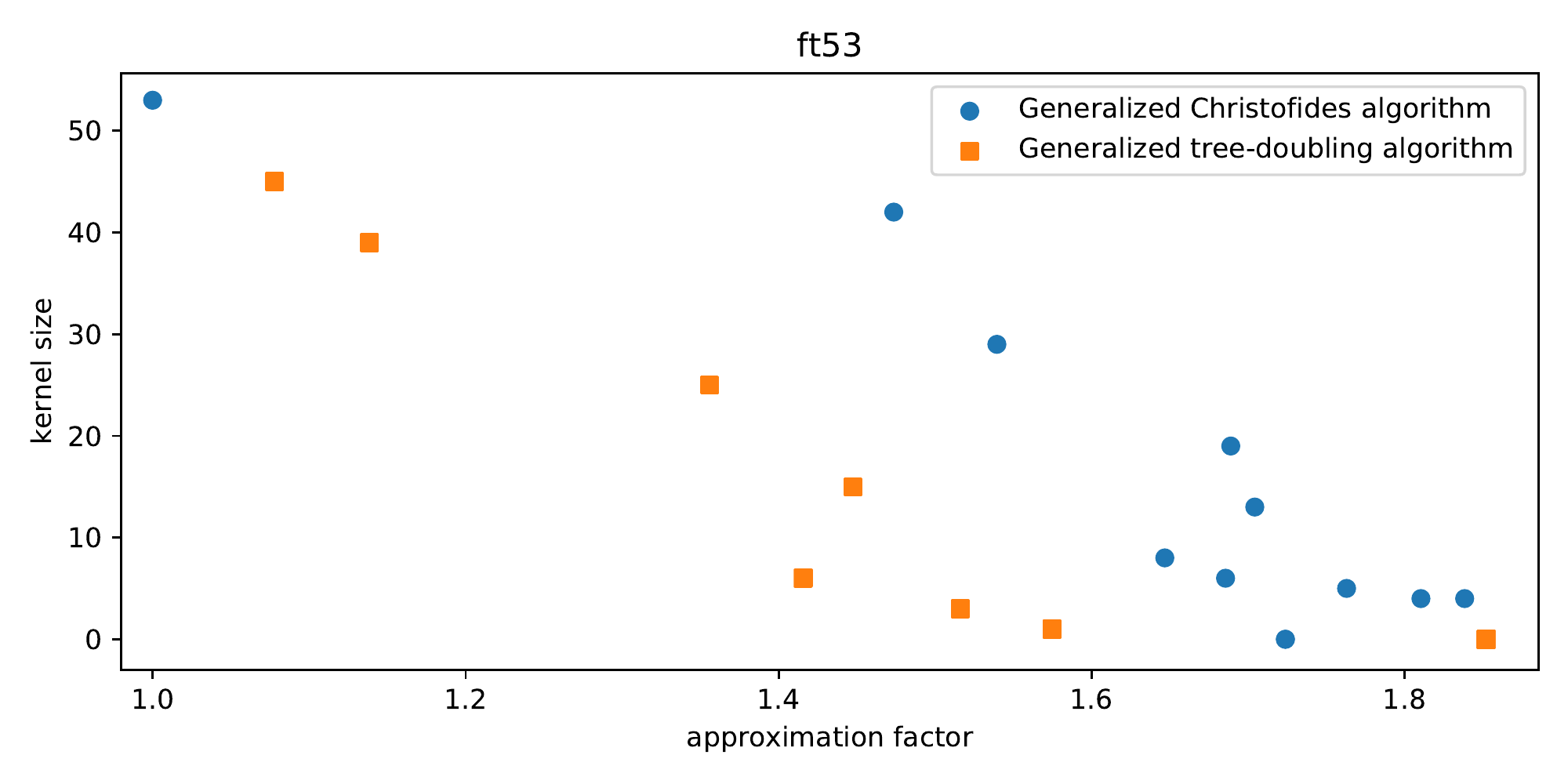}
        \includegraphics[width=0.97\textwidth]{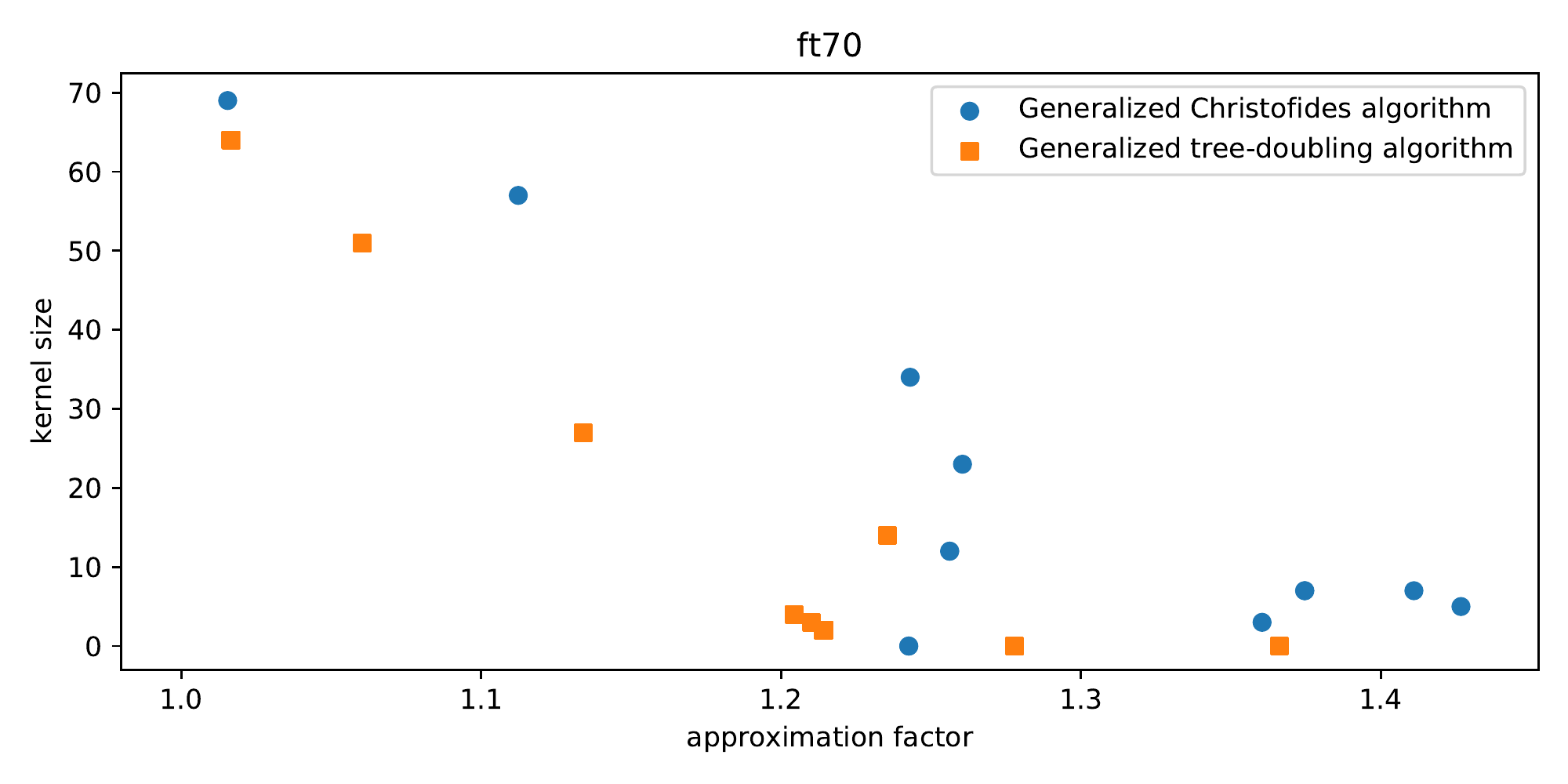}
    \includegraphics[width=0.97\textwidth]{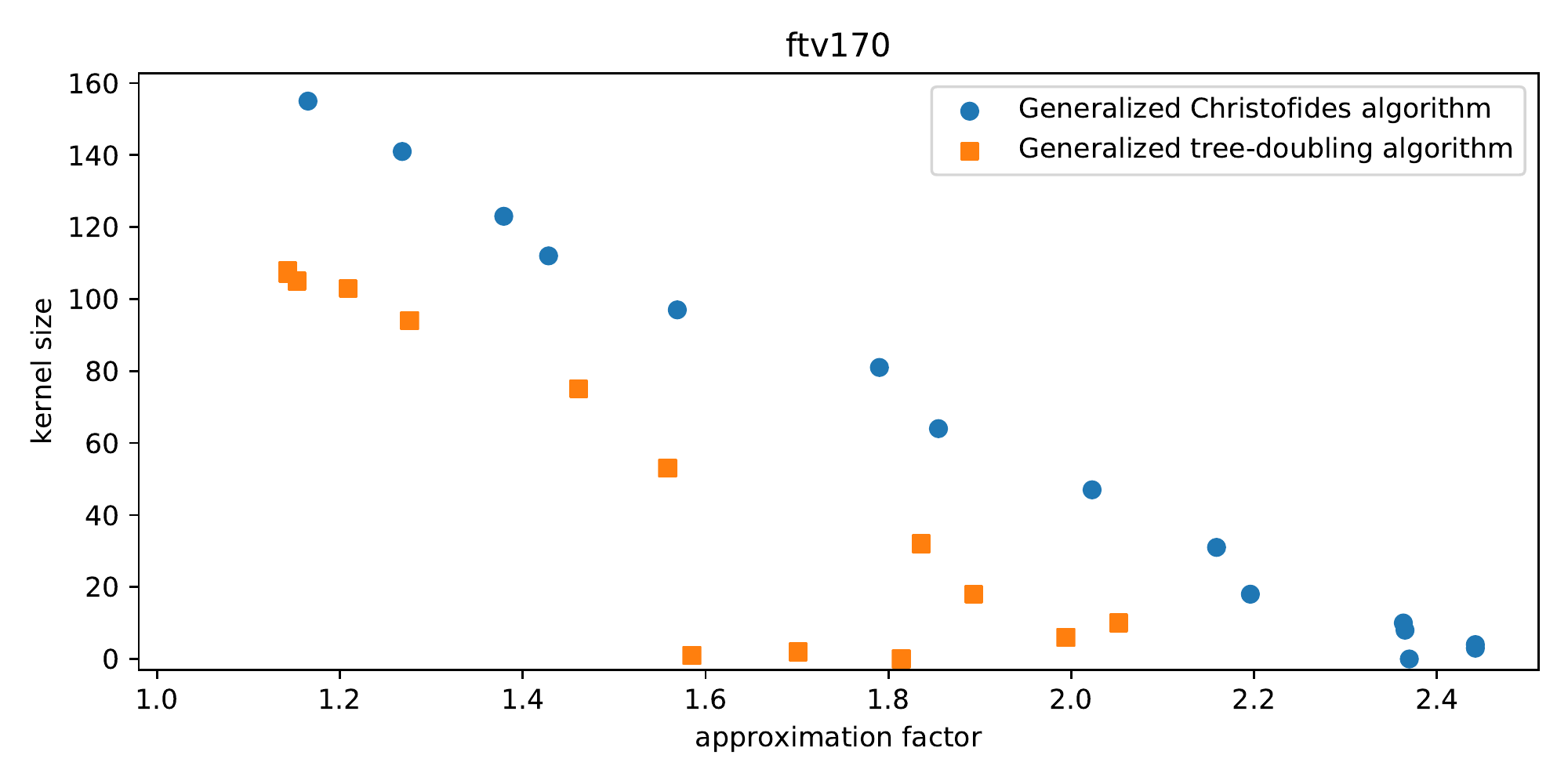}
    \caption{}
    \label{fig:plots1}
\end{figure}

\begin{figure}
    \includegraphics[width=0.97\textwidth]{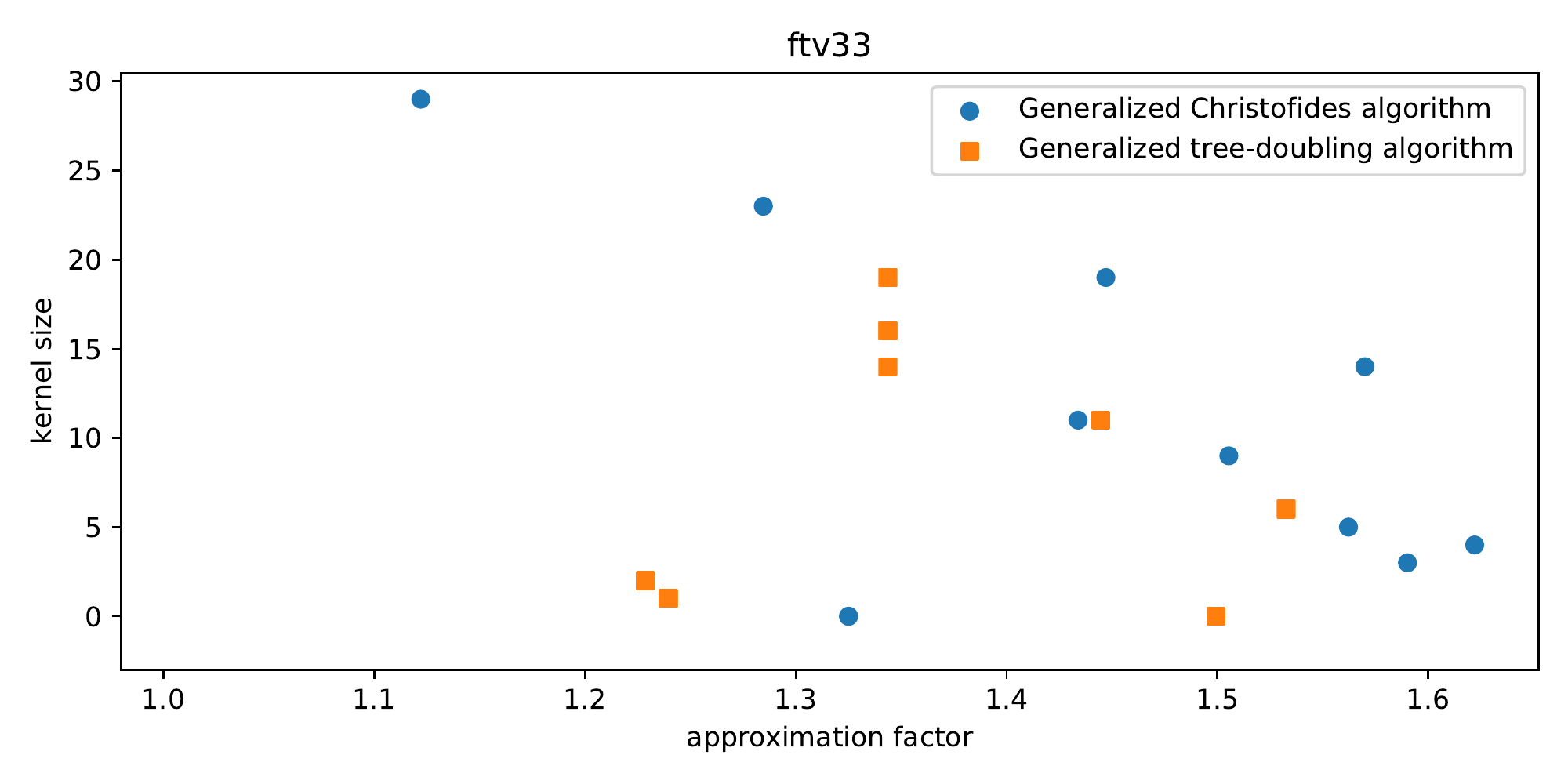}
    \includegraphics[width=0.97\textwidth]{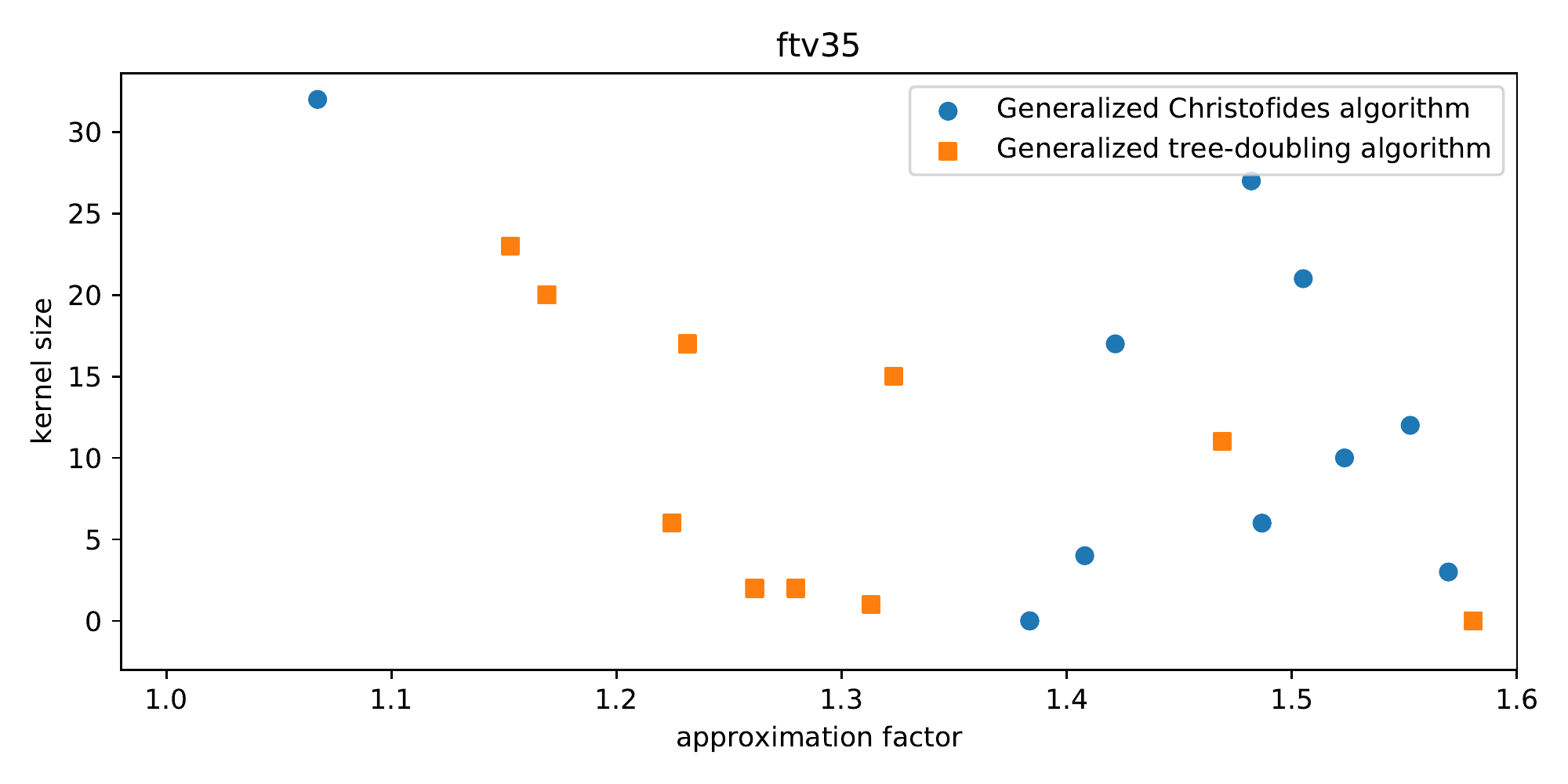}
    \includegraphics[width=0.97\textwidth]{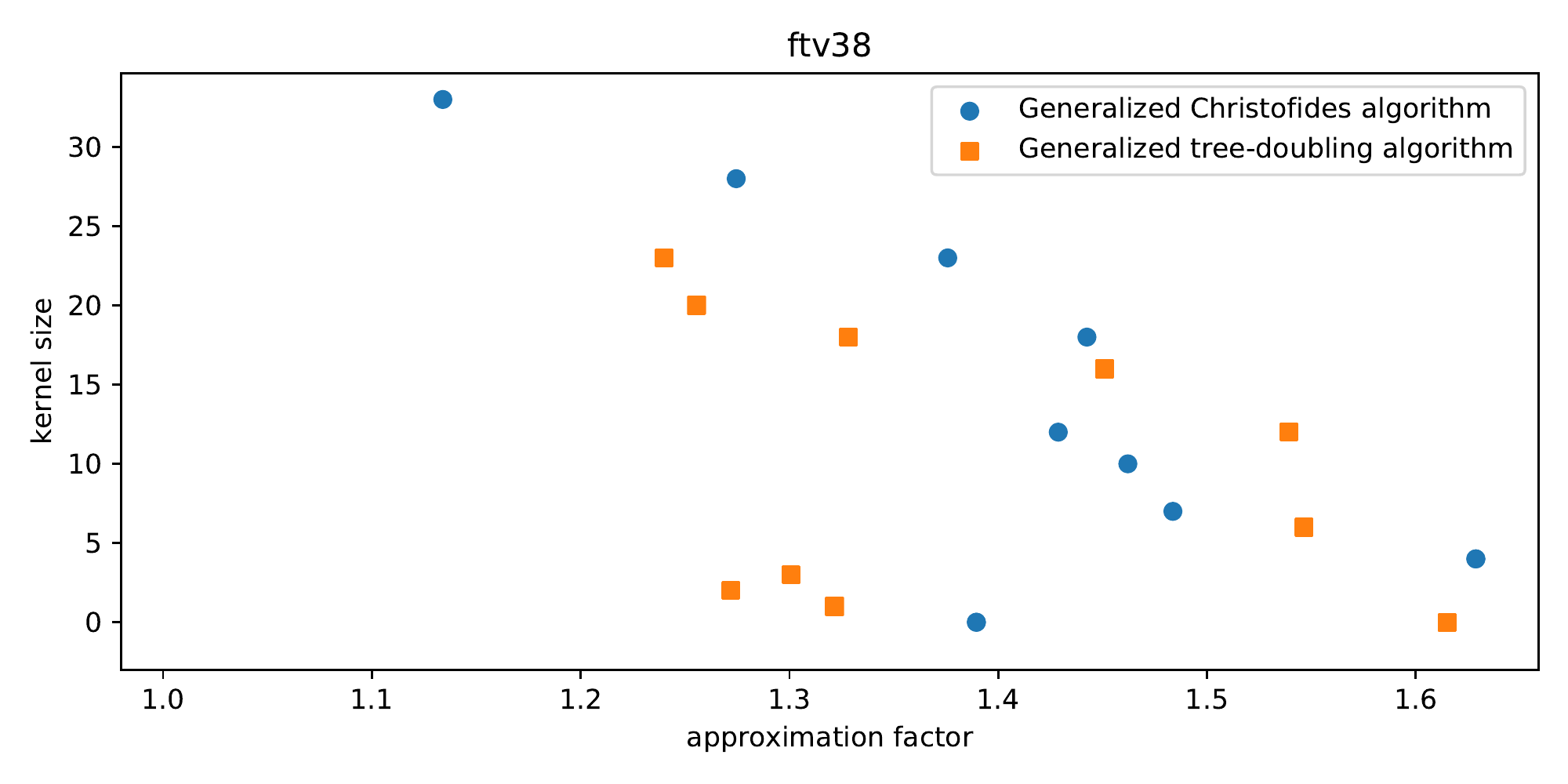}
    \caption{}
    \label{fig:plots2}    
\end{figure}

\begin{figure}
    \includegraphics[width=0.97\textwidth]{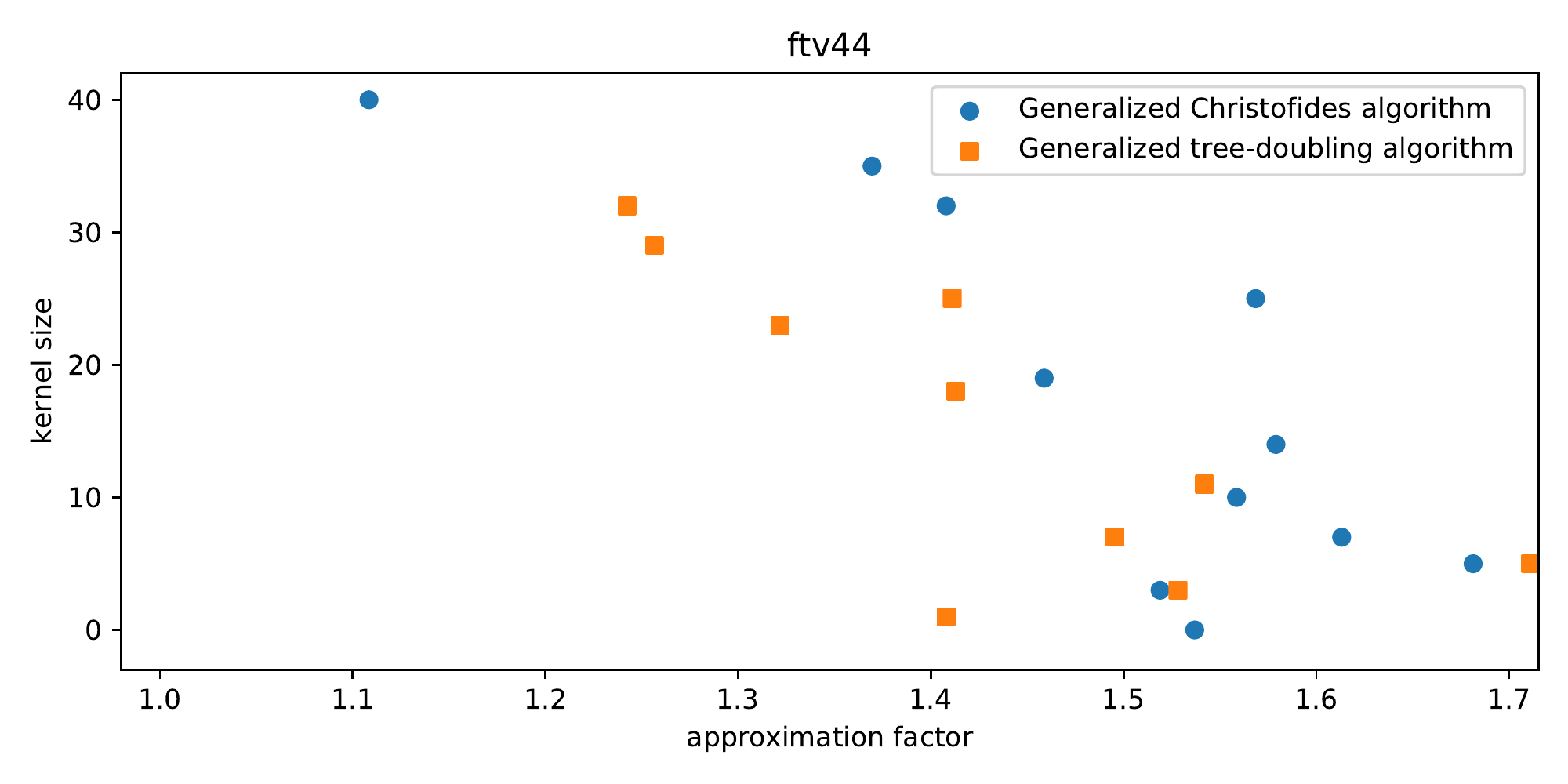}
        \includegraphics[width=0.97\textwidth]{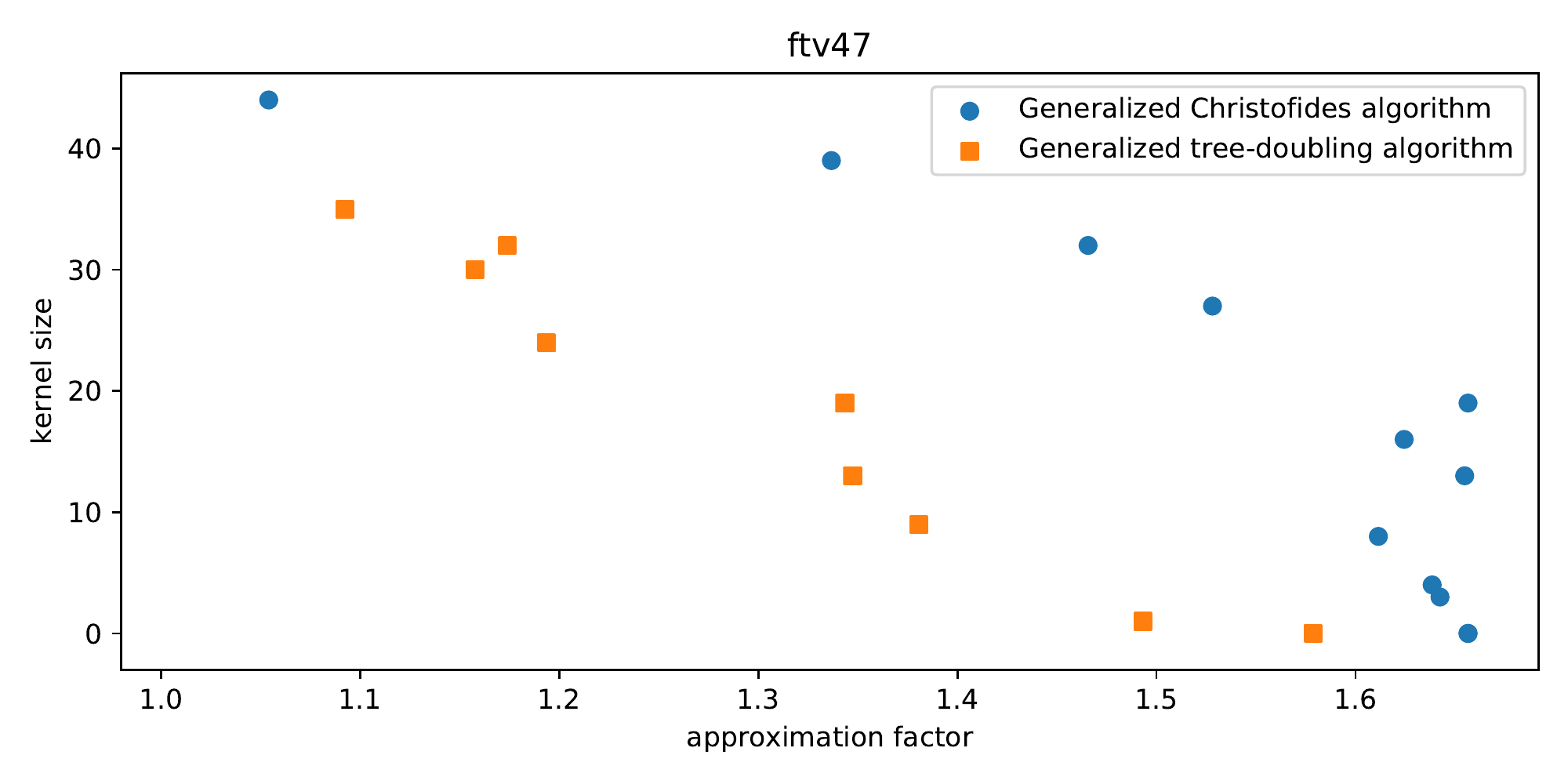}
    \includegraphics[width=0.97\textwidth]{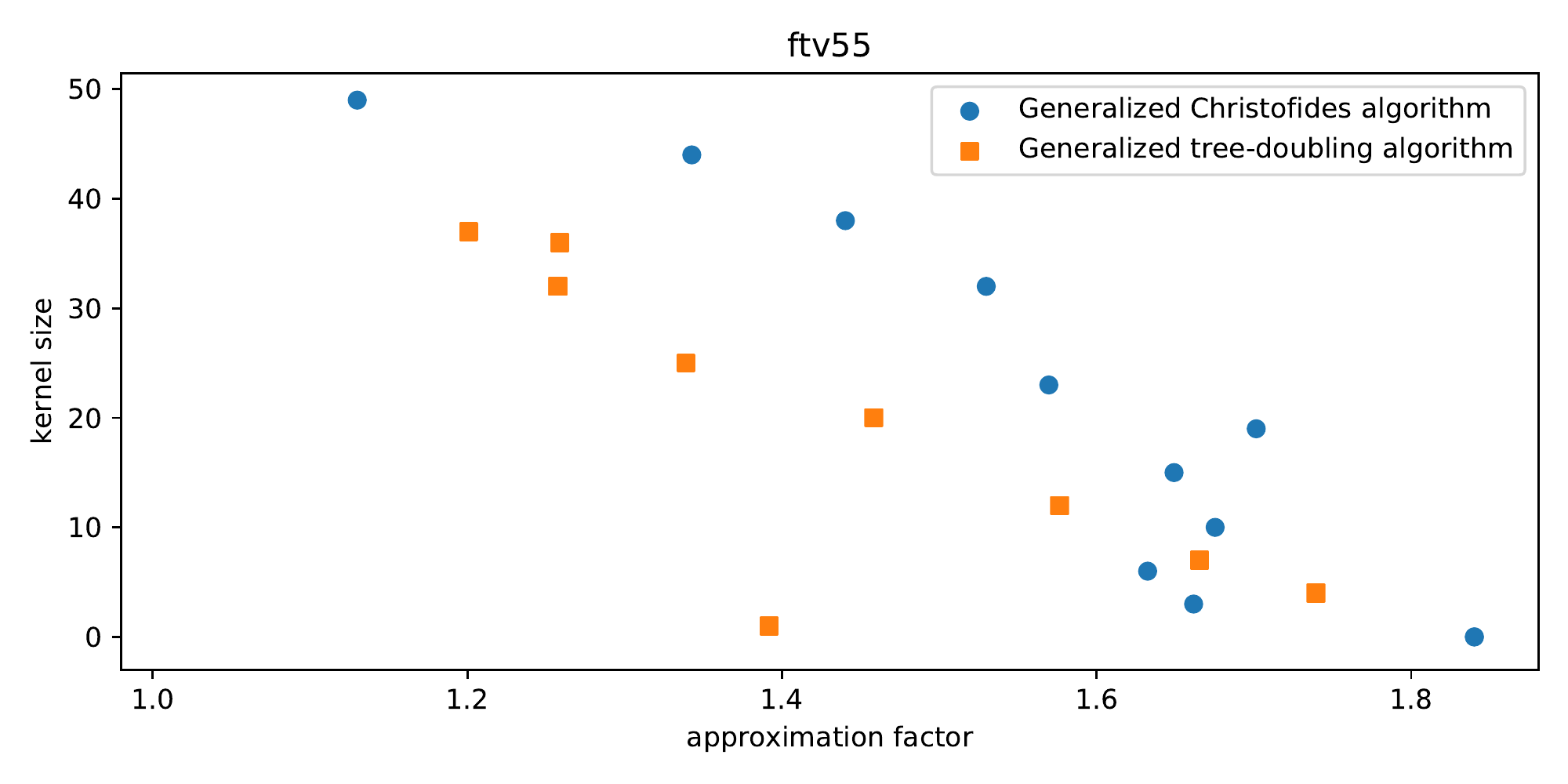}
    \caption{}
    \label{fig:plots3}
\end{figure}

\begin{figure}
    \includegraphics[width=0.97\textwidth]{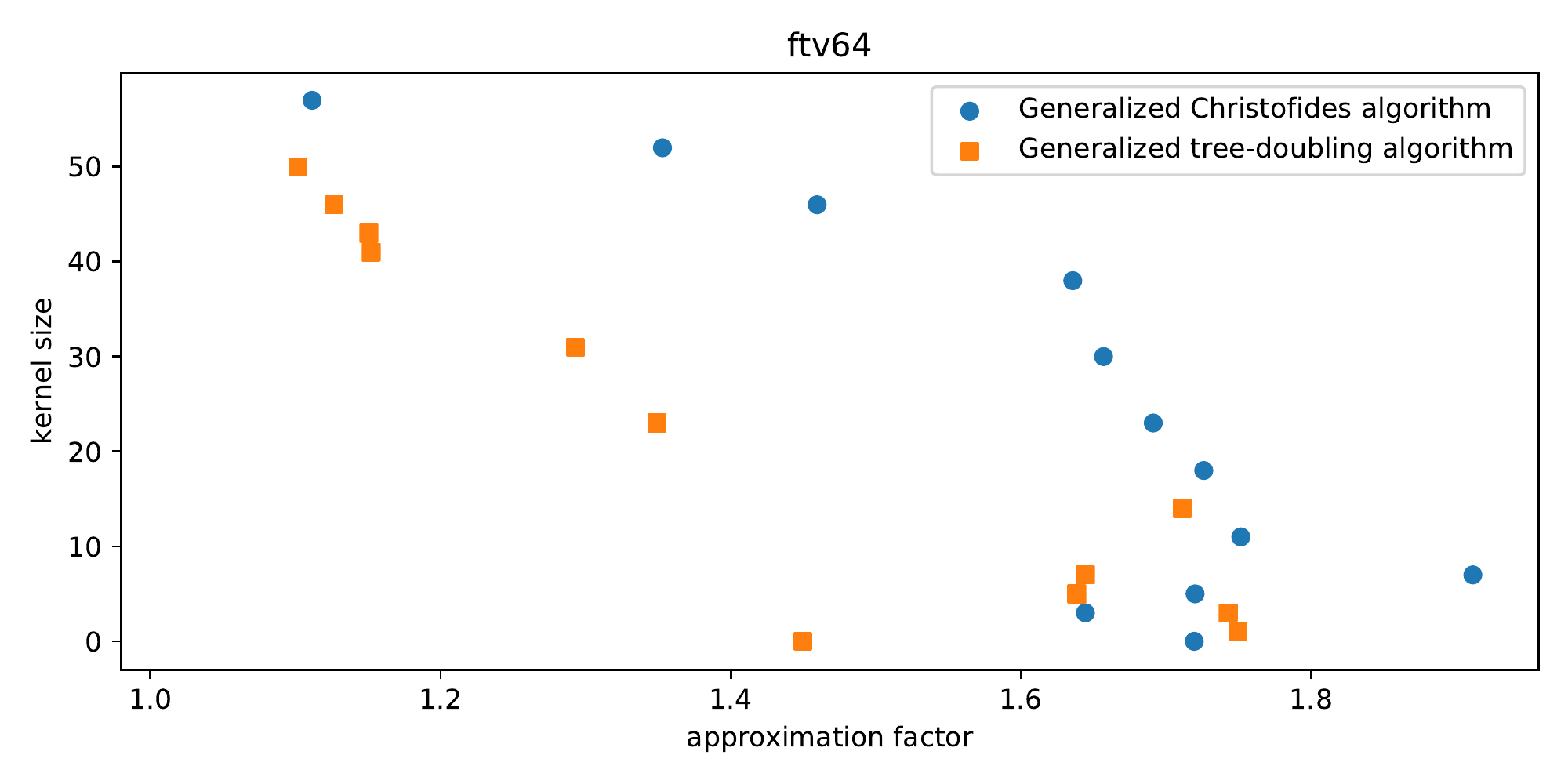}
     \includegraphics[width=0.97\textwidth]{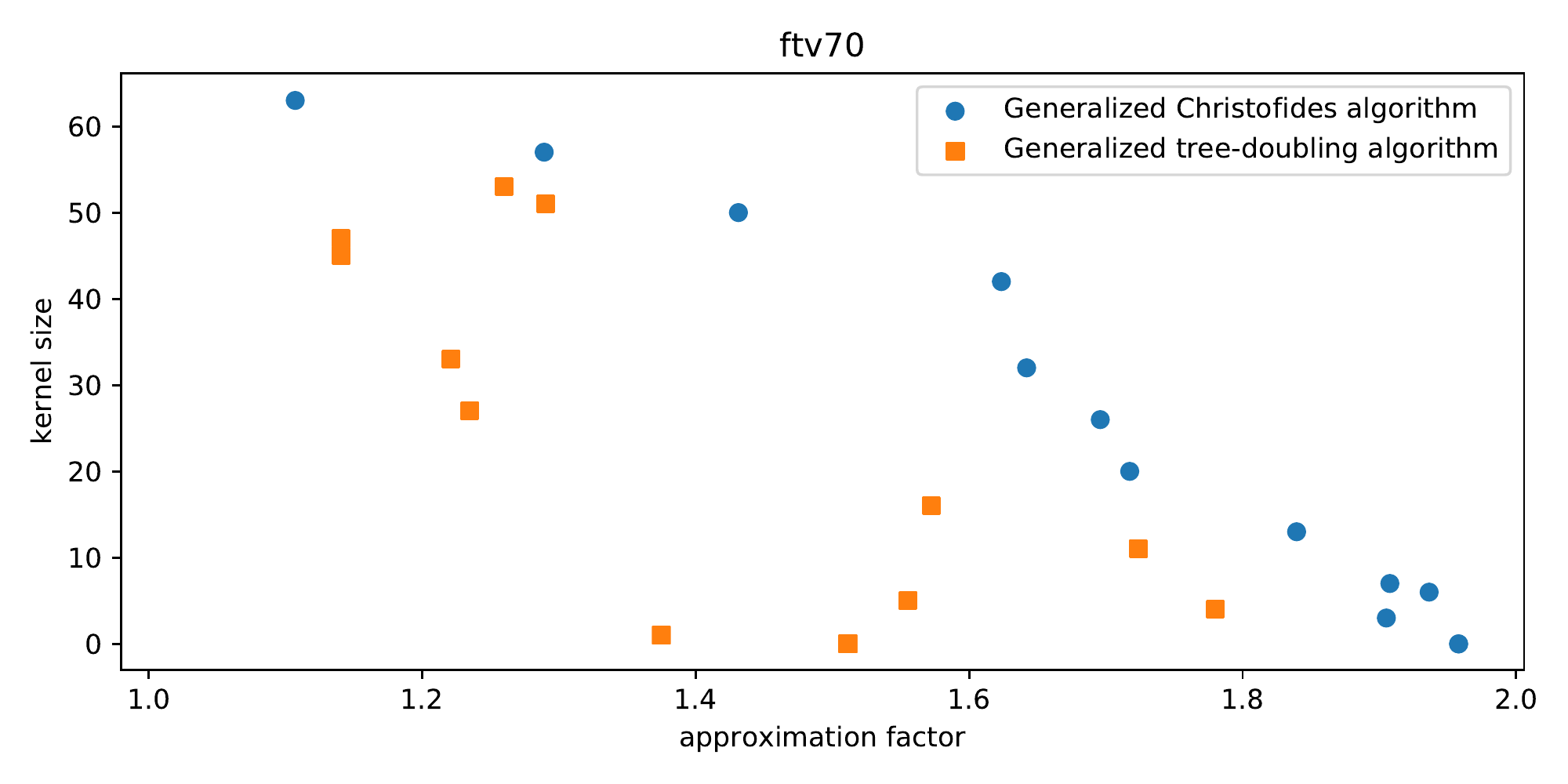}
    \includegraphics[width=0.97\textwidth]{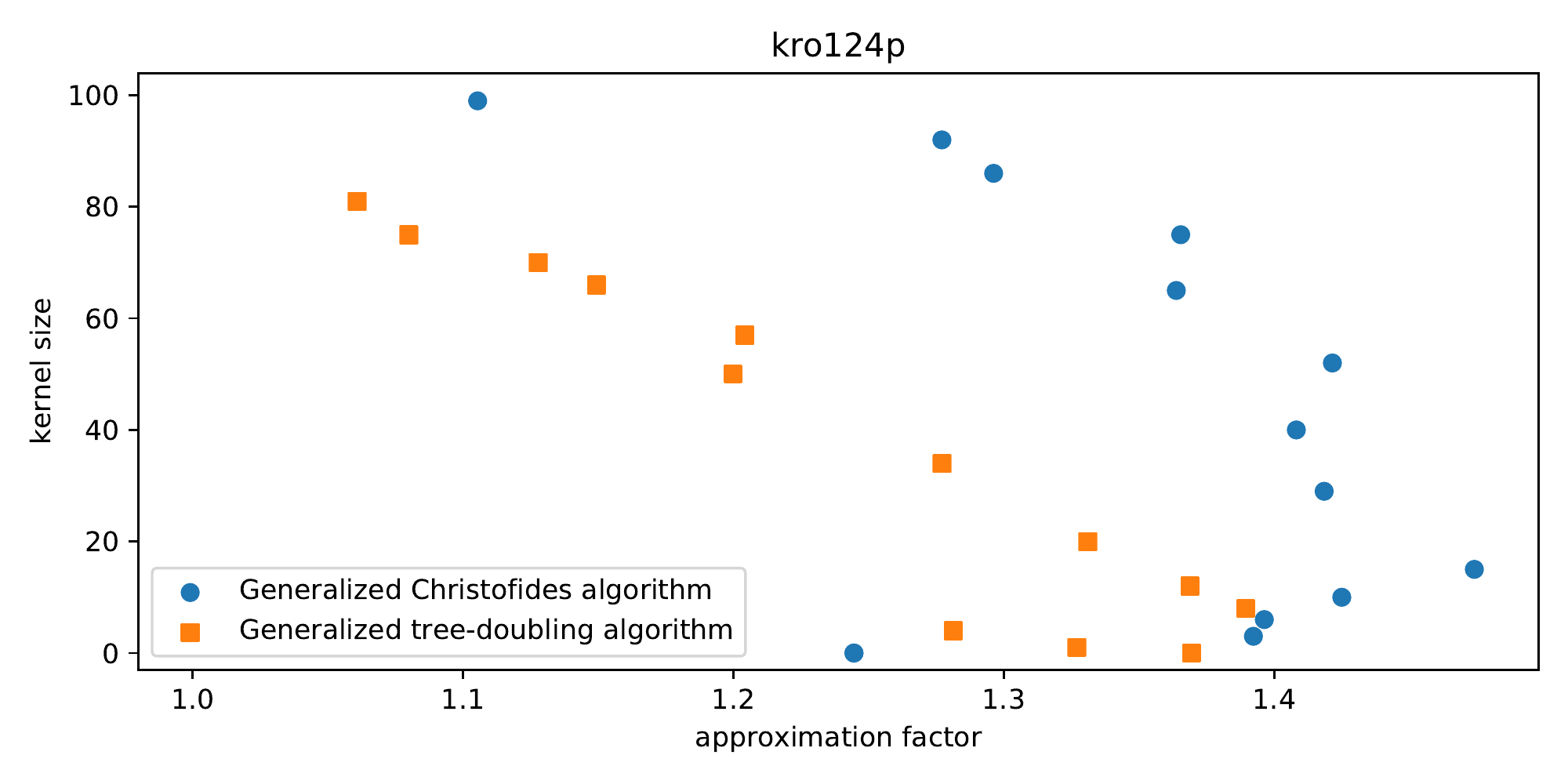}   
    \caption{}
    \label{fig:plots4}
\end{figure}

\begin{figure}
    \includegraphics[width=0.97\textwidth]{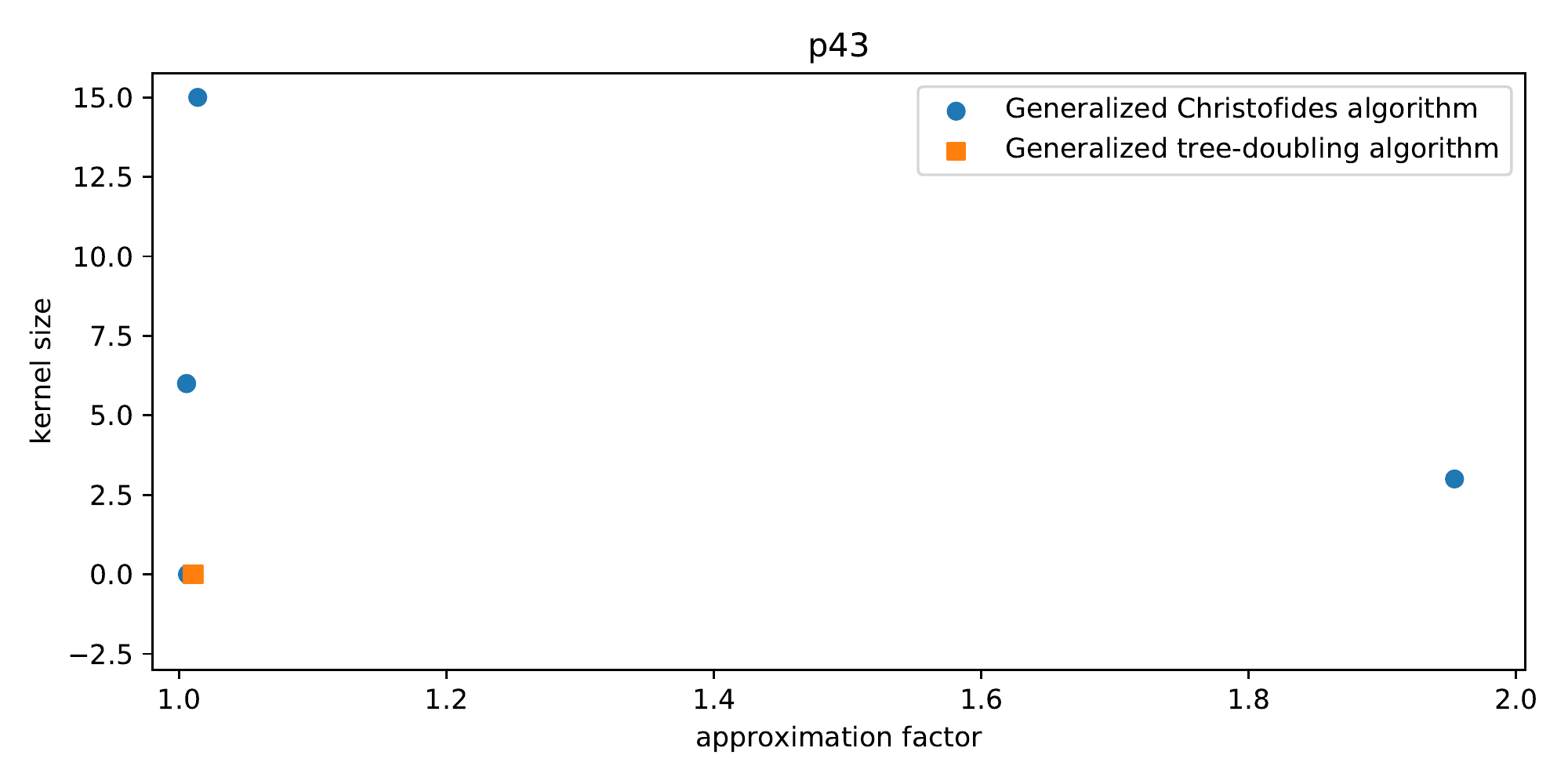}
     \includegraphics[width=0.97\textwidth]{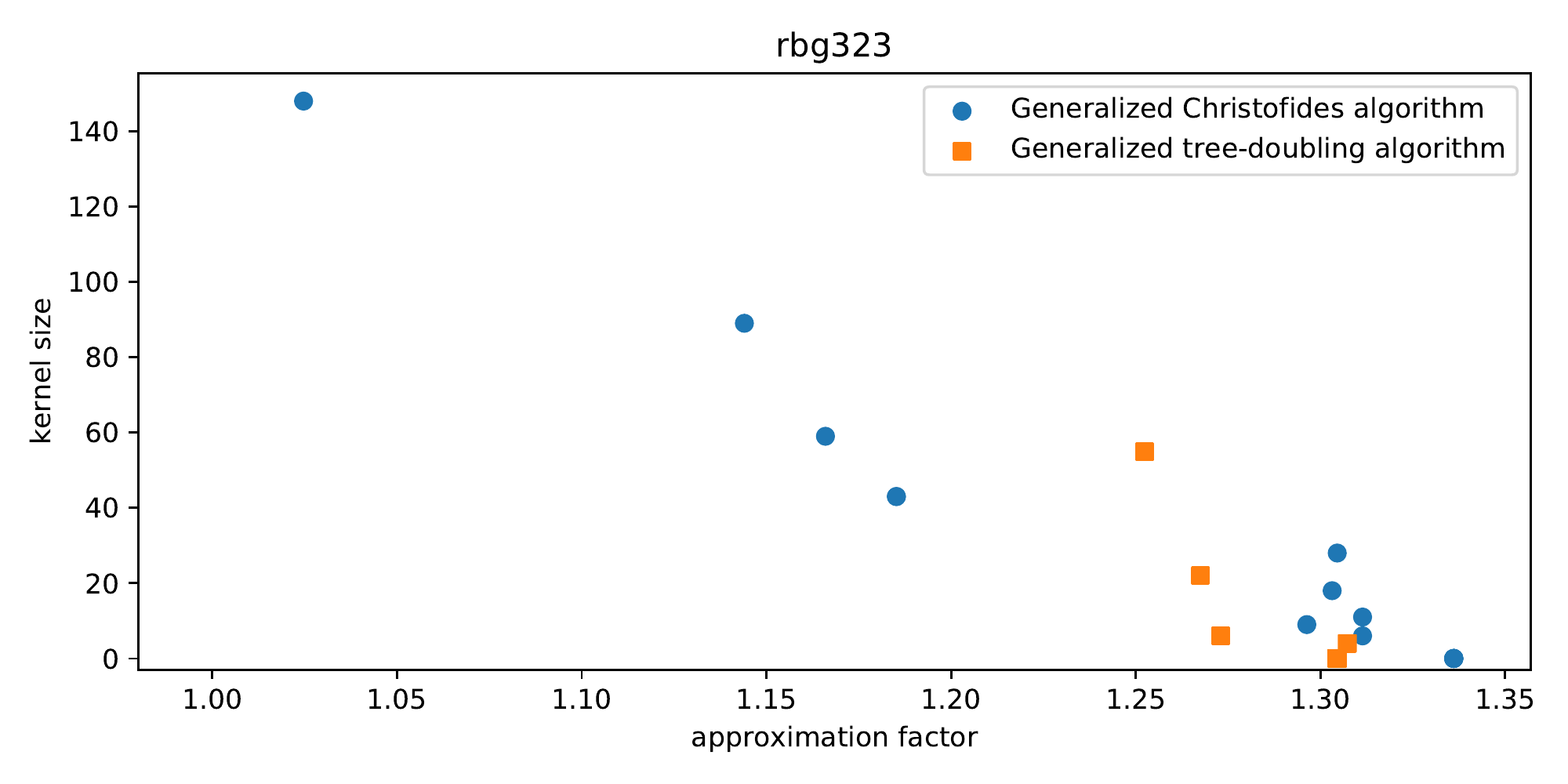}
    \includegraphics[width=0.97\textwidth]{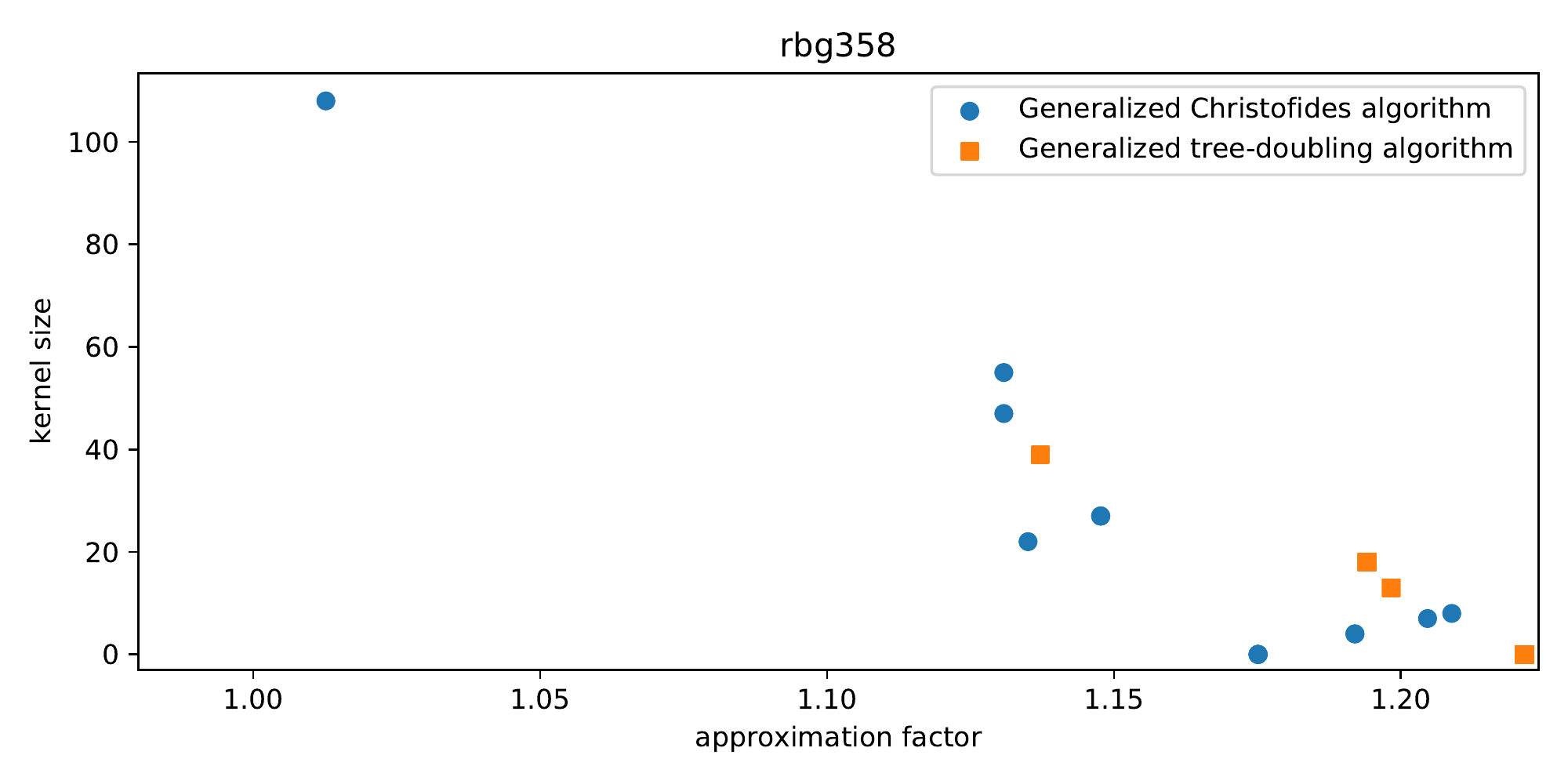}   
    \caption{}
    \label{fig:plots5}
\end{figure}

\begin{figure}

    \includegraphics[width=0.97\textwidth]{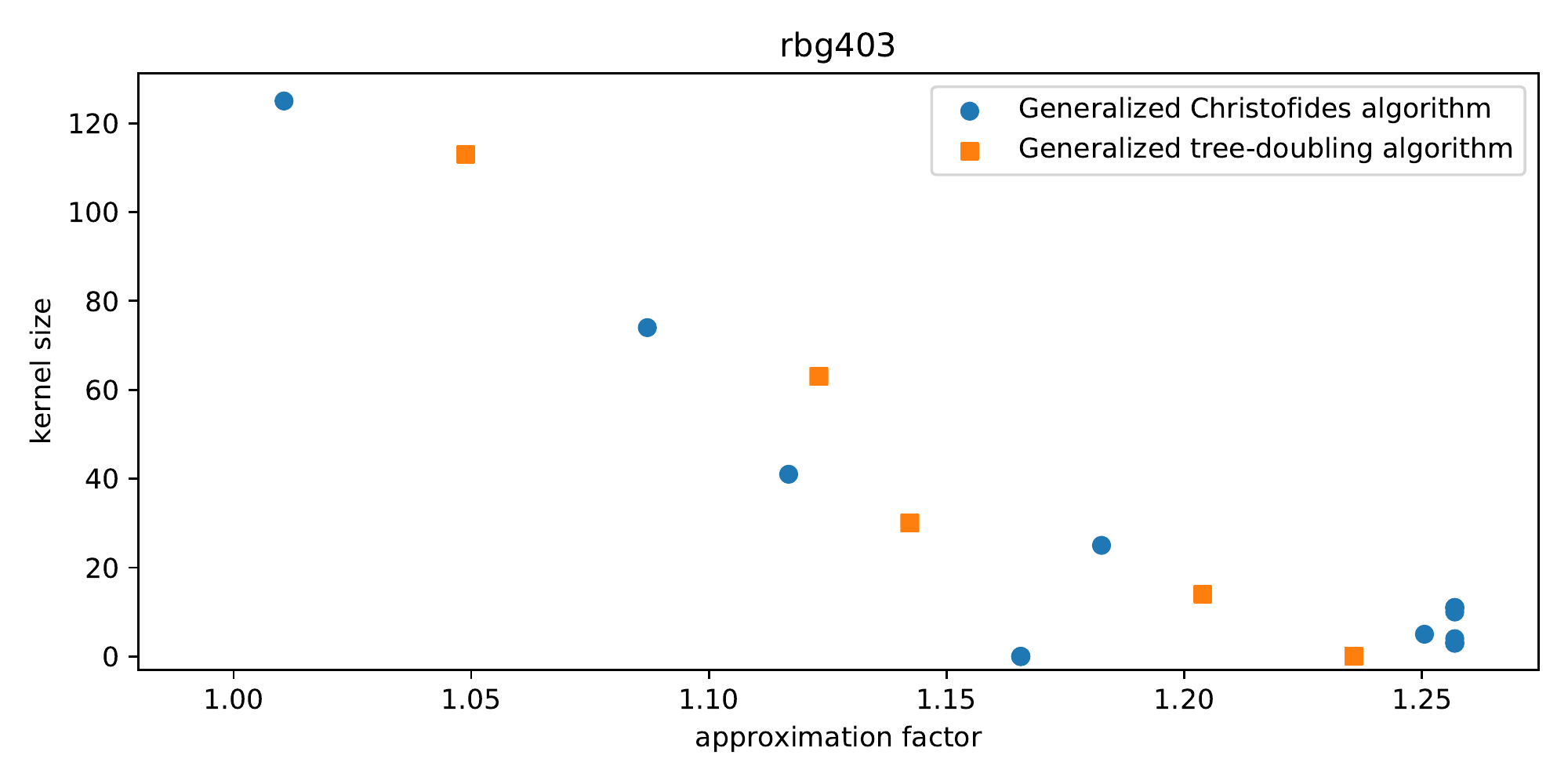}
        \includegraphics[width=0.97\textwidth]{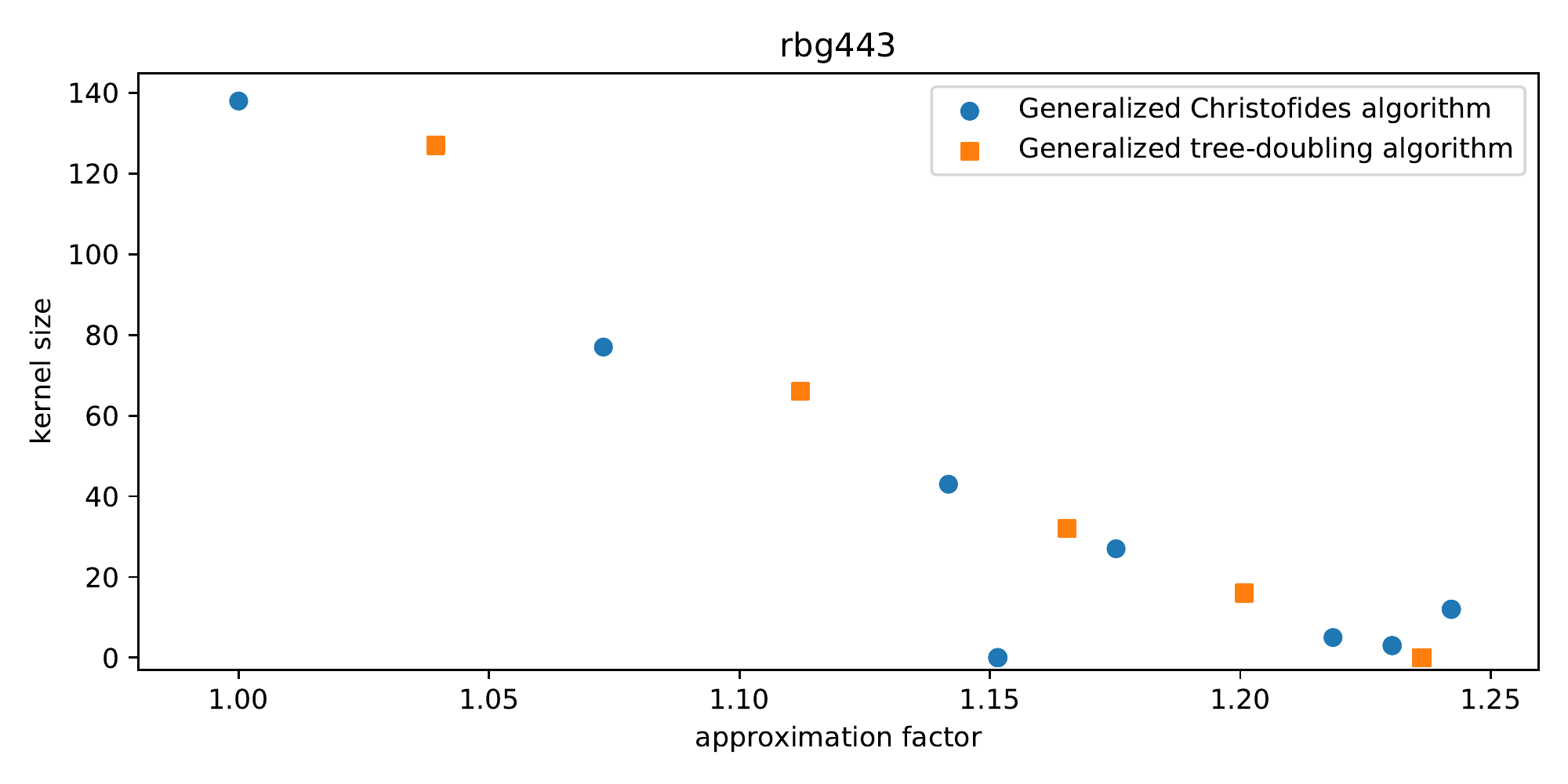}
    \includegraphics[width=0.97\textwidth]{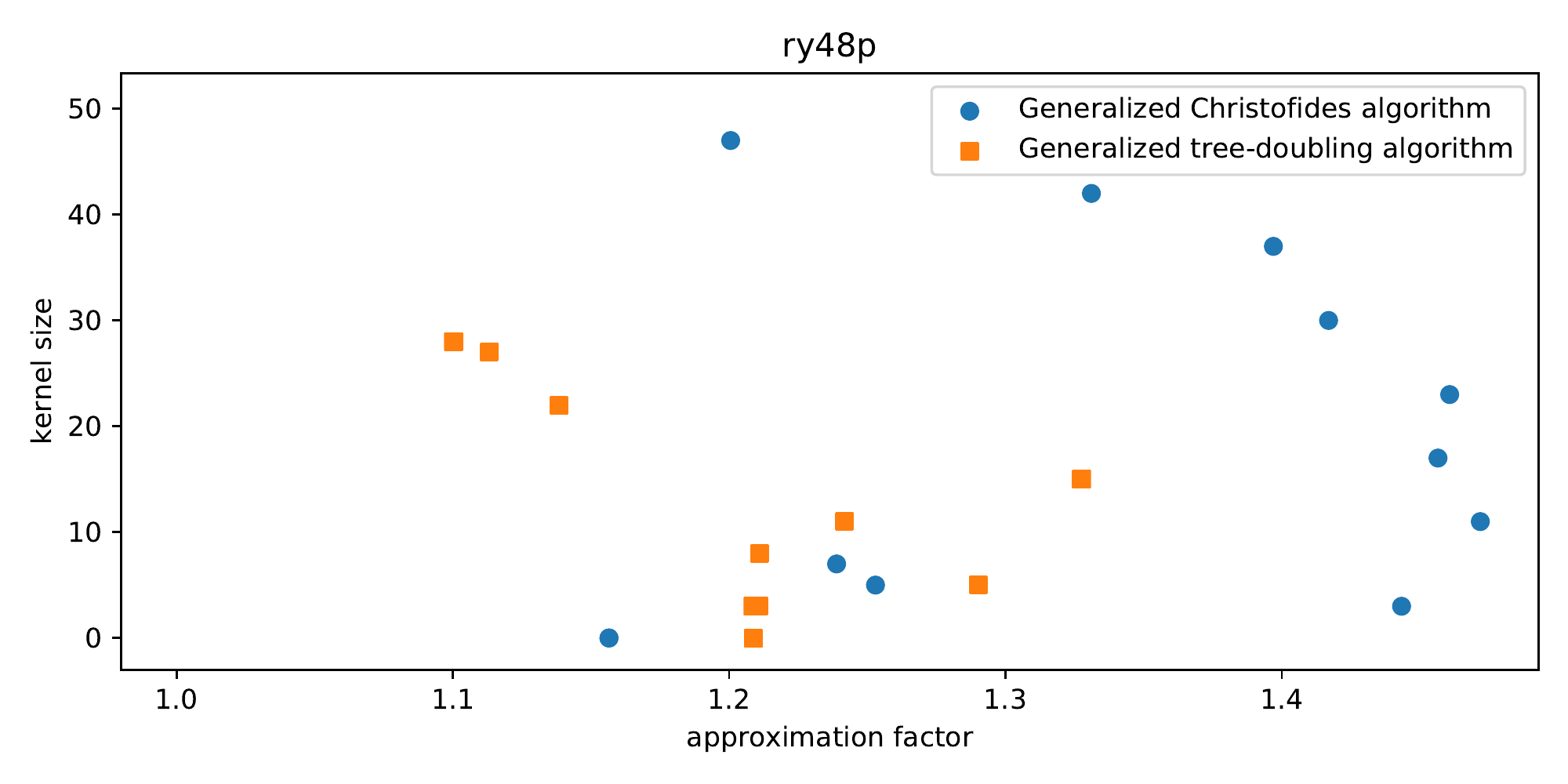}
    \caption{}
    \label{fig:plots6}
\end{figure}

The plots in Figures~\ref{fig:plots1}~to~\ref{fig:plots6} are an alternative visualization of the experimental results shown in Table~\ref{tab:results}.
Each figure visualizes the experiments on a specific graph, and each plotted point represents a single experiment.
Unlike in Section~\ref{sec:experiments}, every new experiment reduces the number of edges treated as asymmetric to a half of the previous experiments, leading to more detailed results. This halving process was continued until the kernel size was zero.
\newpage
\pagestyle{empty}

\end{document}